\documentclass[aps,prx,10pt,twocolumn,superscriptaddress,longbibliography,floatfix]{revtex4-2} % For LaTeX2e
\usepackage{graphicx}
\usepackage{amssymb}
\usepackage{xcolor}
\usepackage[hidelinks,breaklinks]{hyperref}
\usepackage{booktabs}       % professional-quality tables
\usepackage{amssymb,amsfonts}       % blackboard math symbols
\usepackage{mathtools}
\usepackage{nicefrac}       % compact symbols for 1/2, etc.
\usepackage{amsthm}
\usepackage{braket}
\usepackage{bm}

\theoremstyle{plain}

\begin{document}
 
\title{Quantum Classification through Tournament Voting for Robust Single-Shot Inference}
   
\author{Anastasja D. Helgesen}
\email{nathaniel.helgesen@liu.se}
\affiliation{Department of Electrical Engineering, Link\"oping University, Link\"oping, 583 30, Sweden}
\author{Jan-\AA ke Larsson}
\email{jan-ake.larsson@liu.se}
\affiliation{Department of Electrical Engineering, Link\"oping University, Link\"oping, 583 30, Sweden}
\author{Michael Felsberg}
\email{michael.felsberg@liu.se}
\affiliation{Department of Electrical Engineering, Link\"oping University, Link\"oping, 583 30, Sweden}

\begin{abstract}
Quantum machine learning (QML) promises powerful classification capabilities, but suffers from fragile output encodings and high sampling demands---especially in multiclass settings. 
Traditional schemes such as one-hot and binary encoding either produce interpretable outputs too rarely or require many shots to achieve reliable predictions. 
We propose a decision aggregation framework for quantum multiclass classification based on round-robin tournament scoring. 
Each output qubit represents a binary comparison between class pairs, and the final prediction is determined by majority wins. 
This structure improves both the resolvability and accuracy of single-shot predictions, outperforming standard encodings under fixed shot-count conditions. 
Our method retains global entanglement while localizing decision tasks, enabling interpretable inference that remains reliable under intrinsic quantum randomness, without sacrificing expressivity. 
Empirical results show that this approach achieves high accuracy and interpretability under fixed shot-count measurements, suggesting a promising direction for future quantum classifiers. 
\end{abstract}

\maketitle
\section{Introduction}
Quantum machine learning (QML) seeks to harness the unique properties of quantum systems---such as superposition, entanglement, and interference---to perform learning tasks that may be intractable for classical models. 
A central tool in quantum machine learning is the parameterized quantum circuit (PQC), a variational quantum model that applies trainable quantum gates to optimize a task-specific objective function \citep{cerezo_variational_2021,schuld_machine_2021}. 
These circuits are often trained using classical optimization techniques, and their outputs are typically interpreted via expectation values or discretized measurements. 

Despite the theoretical promise of QML, an inherent sampling bottleneck poses a fundamental challenge for scalable quantum inference and will persist as a key consideration for future applications. 
Quantum measurements collapse highly-expressive quantum states into binary outcomes, requiring repeated executions of the circuit to draw samples—or shots—to estimate meaningful statistics \citep{schuld_machine_2021}. 
This limitation is particularly acute in multiclass classification, where the structure of the output encoding plays a critical role. 

In one-hot encoding schemes, the proportion of resolvable outputs---i.e., those that correspond to valid class predictions---vanishes exponentially with the number of classes, making inference increasingly unreliable \citep{chen2024novel}. 
Binary encoding schemes, including standard binary and Gray code, avoid this combinatorial collapse but suffer from a different issue: individual bits are often noisy and weakly correlated with the true class, leading to poor accuracy unless a large number of shots are used \citep{larose2020robust}.

These problems are distinct but linked by a common theme: the difficulty of extracting reliable, discrete decisions from quantum models under limited measurement budgets. 
Therefore, rather than focusing on expectation values or aggregate statistics across many shots, we examine the quality of individual measurement samples—what we call \textit{single-shot inference} (not to be confused with few-shot regimes referring to small training datasets). 

\begin{figure*}[bth]
    \centering

      \includegraphics[width=\linewidth, trim={.75cm .2cm .7cm .15cm},clip]{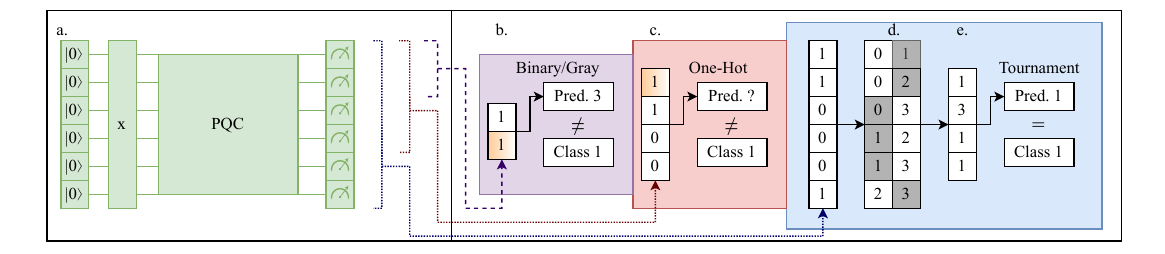}

    \caption{Contrived example illustrating common inference errors across different PQC output encoding strategies and the robustness of our tournament-based approach. 
(a) Standard PQC setup with angular-encoded inputs, a learnable circuit (we test six variants), and Pauli-Z basis measurements.
(b) Binary (similarly Gray code) encoding demonstrating misclassification, predicting the wrong class.
(c) One-hot encoding demonstrating a nonclassification, predicting no class.
(d) Our tournament mapping decomposes multiclass inference into pairwise quantum comparisons, where each output represents a vote between two classes (darker box indicates the chosen class).
(e) Final class is determined by tallying votes across all comparisons.
As the number of classes increases, the tournament structure introduces redundancy that helps mitigate both misclassification and nonclassification, improving single-shot inference reliability.}
\label{fig:training}
\end{figure*}

To evaluate performance under these constraints, we introduce the metric of \textit{shot resolvability}, defined as the probability that a single measurement sample yields a valid and unambiguous class prediction (correct or incorrect).
We address the challenge of shot resolvability by introducing a decision aggregation framework for quantum multiclass classification. 
Rather than relying on global output encodings, our method decomposes the classification task into a series of binary comparisons between class pairs. 
Each comparison is implemented as a binary quantum classifier operating on a shared entangled state. 
The outputs of these classifiers are aggregated using a round-robin tournament structure, where each class competes against every other, and the final prediction is determined by majority wins \citep{tournamentcounts}. 
This approach leverages the statistical robustness of binary decisions and the emergent structure of tournament theory, which ensures that as the number of classes grows, the likelihood of a majority winner (a unique class prediction that wins more pairwise comparisons than any other class) converges to unity while being bound below by $50\%$ \citep{malinovsky2024roundrobintournamentsuniquemaximum}.

Importantly, this framework does not discard the global coherence of the quantum model. 
All classifiers operate within the same entangled quantum state, allowing input information to propagate across the full Hilbert space. The aggregation mechanism simply localizes the decision task, enabling interpretable inference that remains reliable under intrinsic quantum randomness, without sacrificing expressivity. Empirical results show that this method significantly improves accuracy under fixed shot-count regimes, with a particular emphasis on single-shot reliability, outperforming traditional encoding schemes in both sample efficiency and decision consistency.

Our results assume idealized, noiseless conditions to isolate algorithmic behavior from hardware-specific noise. Hardware applicability and noise resilience remain open challenges. Furthermore, our method trades increased, quadratic output-qubit scaling for improved measurement-level resolvability and reduced reliance on repeated sampling which we discuss in Section~\ref{sec:lims}.

This paper makes the following contributions:

\begin{itemize}
    \item We introduce \textit{shot resolvability} as a key metric for evaluating the reliability of single-shot predictions in quantum classifiers, providing a practical lens for assessing inference quality under limited measurement budgets.
    
    \item We propose a novel output encoding for black-box variational quantum classifiers (VQCs) based on round-robin tournament scoring, leveraging the statistical properties of majority-vote tournaments to improve both resolvability and accuracy.

    \item We develop a differentiable training procedure for our tournament-based encoding by embedding pairwise class comparisons into a continuous simplex structure, enabling end-to-end optimization via standard backpropagation.
    
    \item We present a comprehensive empirical evaluation across multiple circuit architectures and datasets, demonstrating that our method consistently outperforms standard global encoding schemes in single-shot regimes 
\end{itemize}

A brief comparison of the three existing encoding strategies and the one proposed in this paper is shown in Table~\ref{tab:encoding_comparison}.

\begin{table*}[t]
    \caption{Comparison of output encoding strategies in terms of validity and accuracy under single-shot and many-shot regimes. }
    \bigskip
    \centering
        \setlength\tabcolsep{2.5mm}
        \begin{tabular}{lccc}
        \toprule
        \textbf{Encoding Method} & \textbf{Resolvability} & \textbf{Accuracy (Single-Shot)} & \textbf{Accuracy (Many-Shot)} \\
        \midrule
        One-Hot         & Low ($\sim K/2^K$) & Moderate--High & High \\
        Binary          & Moderate--High ($\geq1/2$) & Moderate--Low & Moderate--High \\
        Gray code       & Moderate--High ($\geq1/2$) & Moderate--Low & Moderate--High \\
        Tournament (this work) & High ($\to 1$ as $K \to \infty$) & High & High \\
        \bottomrule
        \end{tabular}
    \label{tab:encoding_comparison}
\end{table*}

\section{Related Work}
\label{sec:encodings}

QML has produced a wide range of classification models, including quantum adaptations of support vector machines \citep{rebentrost2014quantum}, convolutional neural networks \citep{Cong_2019, bokhan2022multiclass}, and generative models \citep{benedetti_generative_2019}. Many of these rely on hybrid architectures, where a PQC is embedded within a classical pipeline \citep{chalumuri_hybrid_2021, stein_quclassi_2022, shi_hybrid_2023, liu_hybrid_2021}. While effective in simulation, hybrid models typically depend on expectation values or floating-point outputs, which require extensive sampling. 

Recent work has explored direct multiclass classification using PQCs without hybridization \citep{zhou_multi-classification_2023, hur_quantum_2022, shen_classification_2024}, but these approaches often rely on thresholding or maximum selection over expectation values, which again necessitate high shot counts. Moreover, most prior methods use global output encodings such as one-hot or binary schemes, which suffer from either low resolvability or poor robustness to bit-level noise \citep{chen2024novel, larose2020robust, Di_Matteo_2021}. Some recent efforts have explored alternative encodings such as amplitude-based or angle-based schemes \citep{schuld_circuit-centric_2020}, but these typically require deeper circuits or more complex post-selection.

Most prior work in quantum multiclass classification relies on global output encodings such as one-hot, binary, or Gray code representations. 
One-hot encoding assigns each class to a unique qubit, with the correct class represented by a single qubit in the excited state (e.g., $\ket{1}$) and all others in the ground state ($\ket{0}$). 
This encoding is conceptually simple and widely used in both classical and quantum multiclass classification  \citep{bokhan2022multiclass, dhara2024multi}. 
However, as already mentioned, in quantum settings it suffers from a severe validity bottleneck: only $K$ out of $2^K$ possible bitstrings correspond to valid one-hot outputs, where $K$ is the number of classes. 
Thus, the probability of obtaining a resolvable output from a random measurement decays exponentially as $P_{\text{valid}} =K/2^K$, making inference unreliable under few-shot conditions \citep{chen2024novel}. 

Both binary and Gray code encodings map each class label to a binary representation across $\lceil \log_2 K \rceil$ qubits. 
These encodings are highly efficient in terms of qubit usage and have maximal resolvability: every bitstring corresponds to a class label, modulo padding for non-power-of-two class counts.
However, robustness to sampling variability under few-shot inference is poor.
Individual qubits contribute to multiple bits of the class label, and noise in any bit can lead to misclassification. 
Moreover, binary encoding is sensitive to Hamming distance errors, where small---even single-qubit---perturbations in the bitstring can result in large semantic shifts in class prediction \citep{larose2020robust,Ding_2025}. 

Gray code encoding mitigates some of this sensitivity if class labels have a notion of adjacency, by ensuring adjacent class labels differ by only one bit. 
This reduces the impact of single-bit errors, improving robustness under low-shot conditions.
However its accuracy gains over plain binary encoding under single-shot inference are modest and context-dependent. 
Gray code has been used in quantum classification tasks \citep{Di_Matteo_2021,bokhan2022multiclass}, but these examples still lack the needed semantic structure and remain vulnerable to small permutations in the sampled bitstring.
Gray code has been shown to reduce variance in VQE solutions and improve circuit depth efficiency in Hamiltonian simulations, but its benefits in classification are primarily due to reduced bit-flip sensitivity rather than improved decision structure \citep{Di_Matteo_2021}. 

To improve single-shot reliability, we draw inspiration from tournament theory. 
Round-robin tournaments have long been studied as a framework for pairwise comparison and ranking \citep{zermelo1929berechnung, tournamentcounts}. 
Recent results show that the probability of a majority winner in a random tournament converges to unity as the number of classes increases \citep{malinovsky2024roundrobintournamentsuniquemaximum}. 
This observation will be useful in what follows.

\section{Method}

Motivated by the observations in Section~\ref{sec:encodings}, we propose a tournament-based aggregation of quantum pairwise comparisons to yield robust multiclass predictions. 
Our design in Section~\ref{sec:theory} constitutes a decision aggregation framework in which each output qubit represents a binary comparison between class pairs, and the final prediction is determined by majority wins.
This new design requires a post-processing method used to differentiably train a PQC to output quality round-robin tournament results, which we introduce in Section~\ref{sec:edgetraining}, along with some other training decisions.
After these main contributions, we proceed to test the new framework, using circuit setups and variations discussed in Section~\ref{sec:circuitdesign}, and computational tools found in Section~\ref{sec:tools}.

\subsection{Tournament-based encoding}
\label{sec:theory}
Our proposed encoding frames multiclass classification as a round-robin tournament among class pairs \citep{moon1968topics}. 
Each output bit represents a binary decision between two classes, and the final prediction is determined by majority wins. 
This is known as Copeland's method, and corresponds to an orientation of a complete directed graph over $K$ vertices, with $K(K-1)/2$ pairwise decisions.

At inference time, each measured bitstring is decoded by assigning one vote to the winning class in each pairwise comparison. The predicted class is the class with the largest number of wins. A shot is considered resolvable if this maximum is unique; ties for the largest number of wins are treated as unresolvable. This convention separates prediction validity from correctness and is used consistently in the shot-resolvability metrics reported in Section~\ref{sec:metrics}.

The probability that such a \emph{unique} Copeland-style winner exists in a random tournament tends to 1 as the number of classes increases \cite{malinovsky2024roundrobintournamentsuniquemaximum}. 
Thus, if the number of classes is large, even stochastic or partially incorrect binary decisions still will yield a valid class prediction.
For a quantum classifier, this motivates a decoding rule in which stochastic or partially incorrect pairwise comparisons can still yield a formally valid class prediction with high probability. The accuracy of that prediction, however, depends on the learned pairwise comparisons and on the noise properties of the device or simulator.

Unlike one-hot or binary encodings, tournament-based encoding does not require global agreement across qubits. Each decision is localized, yet the model retains global coherence via shared entanglement. 
This combines locally interpretable binary decisions with a globally entangled model state, allowing the output representation to introduce redundancy without decomposing the PQC into independent classifiers.

\subsection{Training procedures}
\label{sec:edgetraining}

All training is conducted under noiseless simulation using the same PQC architecture described in Section~\ref{sec:circuitdesign}.
This standard practice in quantum machine learning isolates algorithmic behavior from hardware-specific noise and ensures fair comparison across encodings. Our focus is on how each output encoding is interpreted and optimized during training, giving each method the best opportunity to perform under its own assumptions.

All encodings require a continuous representation for gradient-based optimization.
For binary encodings (binary and Gray code), this is straightforward: the model outputs continuous values that can be trained using binary cross-entropy or distance-based losses against a known bitstring target. One-hot encoding, widely used in both classical and quantum classification \citep{bokhan2022multiclass, chen2024novel}, corresponds to moving the center of mass of a probability simplex toward the correct vertex, and is typically trained using cross-entropy loss.

Our tournament-based encoding presents a unique challenge: it produces a vector of binary comparisons between class pairs, many of which are undefined for a given target class. Specifically, only the comparisons involving the true class $c_k$ have a well-defined target; the rest are structurally ambiguous. This makes it inappropriate to apply a bitwise loss across all outputs. To address this, we introduce a novel continuous training method that leverages the geometric correspondence between round-robin tournaments and the edges of a regular simplex. By interpolating each binary comparison along the edge connecting its two associated class vertices, we construct class-specific mass points within the simplex. This allows us to compute distances to the true class vertex and apply a softmax-based loss, analogous to one-hot training.

Formally, the PQC outputs expectation values $\langle Z_ {ij} \rangle$ for each qubit that encodes binary comparisons between class pairs $(c_i, c_j)$ where $i < j$. 

In all main experiments, we use a monotonicity-reversing sigmoid-like tempering function $\phi$ to map Pauli-Z expectation values from $[-1,1]$ to confidence scores in $[0,1]$, where expectation values near $-1$ correspond to higher probability of measuring $1$ and values near $1$ correspond to higher probability of measuring $0$. The tempering function is scaled so that gradients are largest near $\langle Z\rangle=0$, where measurement outcomes are most uncertain, and a designated minimum gradient value is achieved at the extrema, where outcomes are more deterministic. The choice of tempering function and scaling was selected by the ablation study reported in Appendix~\ref{sec:tempering} which determined the error-function with a minimum gradient of 0.01 to work best.

This tempering function is applied to produce confidence scores $e_{ij} = \phi(\langle Z_{ij} \rangle)$.
Each score is used to interpolate between the vertices $v_i$ and $v_j$ of a regular, zero-centered $(K-1)$-simplex:
\begin{equation}
\mathbf{p}_{ij} = \bigl(1 - e_{ij}\bigr) \mathbf{v}_i + e_{ij} \mathbf{v}_j.
\end{equation}
For each class $c_k$, we compute the average of the interpolated points along its incident edges:
\begin{equation}
\mathbf{n}_k = \frac{1}{K - 1} \Bigl(\sum_{j < k} \mathbf{p}_{jk}+\sum_{j > k} \mathbf{p}_{kj}\Bigr).
\end{equation}
This yields a class-specific mass point $\mathbf{n}_k$ within the simplex. We then compute the Euclidean distance between each class's mass point and its corresponding vertex:
\begin{equation}
d_k = \| \mathbf{n}_k - \mathbf{v}_k \|,
\end{equation}
and apply a softmax transformation to the inverted distances to produce class scores:
\begin{equation}
p_k = \frac{\exp(1 - d_k)}{\sum_{j=1}^{K} \exp(1 - d_j)}.
\end{equation}

These scores are used in a symmetric cross-entropy loss:
\begin{equation}
\mathcal{L} = -\sum_{k=1}^{K} \left[ y_k \log(p_k) + (1 - y_k) \log(1 - p_k) \right],
\end{equation}
where $y_k$ is the one-hot target label for class $k$. 
This formulation retains the benefits of softmax normalization while preserving class-specific optimization manifolds. Unlike standard cross-entropy, which only penalizes incorrect predictions, symmetric cross-entropy encourages confident separation between correct and incorrect classes. 
This is particularly beneficial in our setting, where each class is defined by its incident binary comparisons. 
Our observations are consistent with prior work showing that symmetric cross-entropy improves class separation and robustness to sampling variability \citep{wang2019symmetric, das2019separability, huang2020balancedsymmetriccrossentropy}. 

One-hot training uses the same softmax symmetric cross-entropy directly on the activations of the expectation values $\langle Z_i \rangle$ from the PQC for each qubit that encodes a class $i$.
For binary and Gray code encodings, we use the same symmetric cross-entropy formulation, omitting the softmax normalization step, as the targets are bitstrings rather than one-hot vectors.

Each PQC model is trained with a batch size of 32 for 6 epochs using the Adam optimizer with an exponential decay learning rate scheduler \citep{kingma2017adammethodstochasticoptimization}, with a starting learning rate of 0.01, a decay rate of 0.9, and scheduler steps equal to one-tenth of the total training steps. This configuration was selected based on an ablation study in Appendix~\ref{sec:optablate}.

\subsection{Circuit Design}\label{sec:circuitdesign}

To ensure consistency and comparability across encoding methods, all experiments use a shared PQC architecture. We adopt the dual-angle encoding scheme from Ref.~\cite{hur_quantum_2022, munikote2024comparing}, which has demonstrated strong performance in prior work. Input features are encoded using $W = \binom{K}{2}$ qubits, where $K$ is the number of classes. Each qubit receives two features—one via a Pauli-X rotation and one via a Pauli-Y rotation—yielding a total of $2W$ encoded features. Input data is scaled to the range $[-1, 1]$ to ensure unique embeddings, and dimensionality reduction is performed using a reproducible autoencoder with dropout \citep{bishop2006pattern}.

The autoencoder was trained only on the training split for each dataset configuration and then applied to the corresponding validation and test data to avoid test-set leakage. Its latent dimension was set to $2W$ to match the number of input rotations in the PQC. The same preprocessing pipeline, random seeds, and trained autoencoder configurations were used across encoding methods for each matched experimental run.

The main circuit topology is a 2-design qubit ring \citep{cerezo_cost_2021}, where each wire is connected to its two neighbors via alternating layers of computational blocks. 
These blocks consist of parameterized single-qubit rotations and two-qubit controlled operations. We evaluate four well-established block types: CNN7 and CNN8 \citep{sim_expressibility_2019, hur_quantum_2022}, SO(4) and SU(4) \citep{wei_decomposition_2012, vatan2004optimal}.
We also test on two versions of a slightly different multi-qubit entangling setup known as Strongly Entangling Layers (SEL) \citep{schuld_circuit-centric_2020}. 
This setup applies parameterized SU(2) rotations on each individual qubit, then applies a 2-qubit controlled gate to each consecutive pair of qubits. 
We test this setup with both CNOT and controlled-Z gates as the 2-qubit gates. 
Descriptions and diagrams of the six types of blocks are provided in Appendix~\ref{sec:blocks}, and a schematic of the overall setup is shown in Fig.~\ref{fig:training}.
We use four layers of ring blocks or SEL layers in all experiments, though this depth can be adjusted to trade off expressivity and gate cost, as shown in Appendix~\ref{sec:depth}. Importantly, our results are not tied to any specific circuit block or depth—our method operates as a post-processing framework and is compatible with a wide range of architectures.

Measurement strategies differ slightly between encoding methods: the one-hot framework measures~$K$ qubits corresponding to class vertices, binary and Gray code frameworks measure $\lceil \log_2 K \rceil$ qubits, and the tournament framework measures all $W$ qubits. Measuring a subset of wires is standard practice in PQC training \citep{hur_quantum_2022, bokhan2022multiclass, zhou_multi-classification_2023, shen_classification_2024, stein_quclassi_2022}, and has even been linked to improved gradient behavior and reduced barren plateau effects \citep{cerezo_cost_2021, Leone_2024,Cerezo_2025}. All measurements are performed in the Pauli-Z basis.

\subsection{Statistics and reproducibility}
\label{sec:stats_repro}

All encoding methods were evaluated using matched random seeds and matched class subsets to enable direct comparison. The main experiments were performed on two datasets, MNIST Digits and Fashion-MNIST, across five randomly selected class subsets and six circuit architectures. 

All confidence intervals are 95\% nonparametric bootstrap confidence intervals. Bootstrap resampling was performed over matched experimental configurations, where a configuration is defined by dataset, class count, class-subset seed, and circuit architecture. For each bootstrap sample, configurations were sampled with replacement and the mean value of each metric was recomputed. Unless otherwise stated, table entries aggregate over both datasets, five class-subset seeds, and six circuit architectures for each value of $K$.

No formal null-hypothesis significance tests were used in the main analysis. Comparisons are instead reported using descriptive statistics over matched experimental runs. 
Raw bitstring shot accuracy, resolvable-shot accuracy, resolvability ratio are defined in Section~\ref{sec:metrics}. 
The resolvability ratio was estimated by sampling until 100 resolvable shots were obtained and dividing 100 by the total number of measurement shots required. Random seeds, class subsets, preprocessing settings, trained-model outputs, and result-generation scripts are provided with the code release.

\subsection{Computational tools}
\label{sec:tools}

All experiments were performed using JAX~\citep{jax2018github} and PennyLane~\citep{bergholm2022pennylaneautomaticdifferentiationhybrid}. 
JAX was used for automatic differentiation, just-in-time compilation, and vectorized execution of batched parameterized quantum circuit evaluations. 
PennyLane was used as the differentiable quantum programming framework for circuit construction, simulation, measurement, and integration with JAX.

The main experiments were performed using noiseless CPU simulation. 
The full experimental suite required approximately 100 kCPU-hours on two Intel Xeon Gold 6130 processors, with an additional approximately 300 kCPU-hours used for ablation studies. 
Because the primary contributions of this work concern output encoding and post-processing, the main experiments are designed to isolate algorithmic behaviour from hardware-specific noise. Supplementary noise-model experiments, described in Appendix~\ref{sec:noise}, evaluate the relative behaviour of selected encodings under retired IBM backend noise models.

In practice, the reported simulations remain feasible for small numbers of classes, here $K \leq 6$, and shallow circuits, here at most four circuit layers. 
These regimes reflect the intended scope of the study: evaluating encoding strategies under strict measurement constraints rather than optimizing for immediate hardware execution. 
The code and data used to reproduce the experiments are described in the Code availability statement and Data availability statement, respectively.

\section{Results}
\label{sec:results}

\begin{table*}[t]
\scriptsize
    \centering

    \caption{Main empirical metrics aggregated by class count and encoding. The table reports resolvability ratio $R$, shot-level resolvable accuracy $A_R^{\mathrm{shot}}$, raw bitstring shot accuracy $A_b^{\mathrm{shot}}$, and continuous top-1 simulation accuracy $T$. The first three metrics evaluate discrete measurement outcomes: $R$ measures how often a shot yields a valid prediction, $A_R^{\mathrm{shot}}$ measures correctness conditional on validity, and $A_b^{\mathrm{shot}}$ counts unresolvable shots as incorrect. The metric $T$ reports accuracy under continuous pre-measurement decoding. Values are means with 95\% bootstrap confidence intervals $[CI^{\downarrow},CI^{\uparrow}]$ over matched experimental configurations. Larger values are better for all metrics. The first three metrics are visualized in Fig.~\ref{fig:metricsplots}; full dataset- and circuit-resolved summaries are provided in Appendix~\ref{sec:fulldata}.}
    \vskip .15in
    \centering
    \setlength\tabcolsep{3.2mm}
    \begin{tabular}{cccccc}
        \toprule
        K & Method & $R$ \;\;[$CI^{\downarrow}$, $CI^{\uparrow}$] \% & 
                     $A_R^{\mathrm{shot}}$ [$CI^{\downarrow}$, $CI^{\uparrow}$] \% & 
                     $A_b^{\mathrm{shot}}$ [$CI^{\downarrow}$, $CI^{\uparrow}$] \% & 
                     $T$\;\;[$CI^{\downarrow}$, $CI^{\uparrow}$] \% \\
        \midrule
 3  & Tournament &  $95.32$\;\;$[94.13, 96.41]$   &  $59.36$\;\;$[57.65, 61.12]$   &  $56.83$\;\;$[55.09, 58.63]$   &  $74.78$\;\;$[72.57, 76.94]$  \\
 3  &  One-hot   &  $56.99$\;\;$[53.19, 60.61]$   &  $60.57$\;\;$[58.49, 62.65]$   &  $40.20$\;\;$[37.49, 42.94]$   &  $74.13$\;\;$[71.83, 76.33]$  \\
 3  &   Binary   &  $91.43$\;\;$[90.22, 92.56]$   &  $55.54$\;\;$[53.56, 57.53]$   &  $51.34$\;\;$[49.47, 53.22]$   &  $73.79$\;\;$[71.07, 76.32]$  \\
 3  &    Gray    &  $92.93$\;\;$[91.81, 94.07]$   &  $55.65$\;\;$[53.81, 57.52]$   &  $52.21$\;\;$[50.21, 54.29]$   &  $75.07$\;\;$[72.78, 77.32]$  \\
\midrule
 4  & Tournament &  $73.78$\;\;$[71.33, 76.15]$   &  $55.41$\;\;$[53.58, 57.25]$   &  $42.42$\;\;$[40.12, 44.71]$   &  $82.02$\;\;$[80.13, 83.85]$  \\
 4  &  One-hot   &  $38.65$\;\;$[35.93, 41.31]$   &  $57.71$\;\;$[55.48, 59.90]$   &  $26.21$\;\;$[23.89, 28.50]$   &  $81.45$\;\;$[79.23, 83.59]$  \\
 4  &   Binary   &  $100.0$\;\;$[100.0, 100.0]$   &  $47.36$\;\;$[45.60, 49.10]$   &  $47.36$\;\;$[45.60, 49.10]$   &  $81.65$\;\;$[79.25, 83.94]$  \\
 4  &    Gray    &  $100.0$\;\;$[100.0, 100.0]$   &  $25.14$\;\;$[24.73, 25.53]$   &  $25.14$\;\;$[24.73, 25.53]$   &  $25.94$\;\;$[25.17, 26.68]$  \\
\midrule
 5  & Tournament &  $68.74$\;\;$[67.15, 70.29]$   &  $47.81$\;\;$[45.35, 50.14]$   &  $34.17$\;\;$[31.84, 36.41]$   &  $74.81$\;\;$[72.40, 77.16]$  \\
 5  &  One-hot   &  $24.45$\;\;$[22.75, 26.16]$   &  $50.22$\;\;$[47.63, 52.71]$   &  $15.73$\;\;$[13.99, 17.46]$   &  $74.55$\;\;$[72.16, 76.92]$  \\
 5  &   Binary   &  $80.10$\;\;$[78.83, 81.35]$   &  $40.53$\;\;$[38.54, 42.53]$   &  $33.20$\;\;$[31.16, 35.21]$   &  $72.28$\;\;$[69.44, 75.10]$  \\
 5  &    Gray    &  $80.08$\;\;$[78.78, 81.39]$   &  $41.19$\;\;$[39.20, 43.14]$   &  $33.71$\;\;$[31.66, 35.71]$   &  $73.21$\;\;$[70.41, 75.89]$  \\
\midrule
 6  & Tournament &  $67.65$\;\;$[66.77, 68.53]$   &  $41.87$\;\;$[38.78, 44.85]$   &  $29.34$\;\;$[26.85, 31.77]$   &  $68.06$\;\;$[64.80, 71.21]$  \\
 6  &  One-hot   &  $15.57$\;\;$[14.67, 16.51]$   &  $41.25$\;\;$[38.36, 44.10]$   &    $8.25$\;\;$[7.10, 9.44]$    &  $66.49$\;\;$[63.00, 69.91]$  \\
 6  &   Binary   &  $83.63$\;\;$[82.86, 84.38]$   &  $31.01$\;\;$[29.25, 32.79]$   &  $26.36$\;\;$[24.68, 28.05]$   &  $61.44$\;\;$[57.88, 64.96]$  \\
 6  &    Gray    &  $83.98$\;\;$[83.16, 84.76]$   &  $31.51$\;\;$[29.76, 33.27]$   &  $26.94$\;\;$[25.25, 28.63]$   &  $62.45$\;\;$[58.66, 66.10]$  \\
        \bottomrule
    \end{tabular}\medskip
    \label{tab:main_results}
\end{table*}

We evaluate tournament decoding against one-hot, binary, and Gray-code encodings under matched experimental conditions. All methods are trained and evaluated using the same class subsets, random seeds, and circuit architectures. Experiments use MNIST Digits and Fashion-MNIST with five random class subsets for each $K$ and six circuit architectures per configuration. Robustness to class imbalance and overlapping classes is discussed in Section~\ref{sec:lims}.

\subsection{Metrics}
\label{sec:metrics}

We report metrics designed to separate validity, conditional correctness, and fixed-shot performance under measurement. The shot-level resolvable accuracy $A_R^{\mathrm{shot}}$ is the fraction of resolvable measurement shots that yield the correct class label. The resolvability ratio $R$ is the fraction of measurement shots that are resolvable. We estimate $R$ by sampling until 100 resolvable shots are obtained and dividing 100 by the total number of shots required.

To assess fixed-shot performance directly, we report the bitstring shot accuracy $A_b^{\mathrm{shot}}$, computed over a fixed number of measurement shots per test sample with unresolvable predictions counted as incorrect. This metric estimates the probability that a single measured bitstring yields the correct class prediction and serves as our primary measurement-level accuracy metric.

For comparison, we also report the top-1 simulation accuracy $T$, obtained by selecting the class with the largest continuous pre-measurement score. Unlike the shot-based metrics, $T$ does not include sampling-induced invalidity and therefore helps separate training performance from discrete measurement reliability.

All confidence intervals are 95\% nonparametric bootstrap confidence intervals. Bootstrap resampling was performed over matched experimental configurations, where a configuration is defined by dataset, class count, class-subset seed, and circuit architecture. For each bootstrap sample, configurations were sampled with replacement and the mean value of each metric was recomputed.

\subsection{Resolvability}
\label{sec:resolvable}

\begin{figure*}[t]
    \centering
\begin{minipage}{.49\linewidth}
    \includegraphics[width=\linewidth,trim={1.5mm 1.5mm 1.5mm 1.5mm},clip]{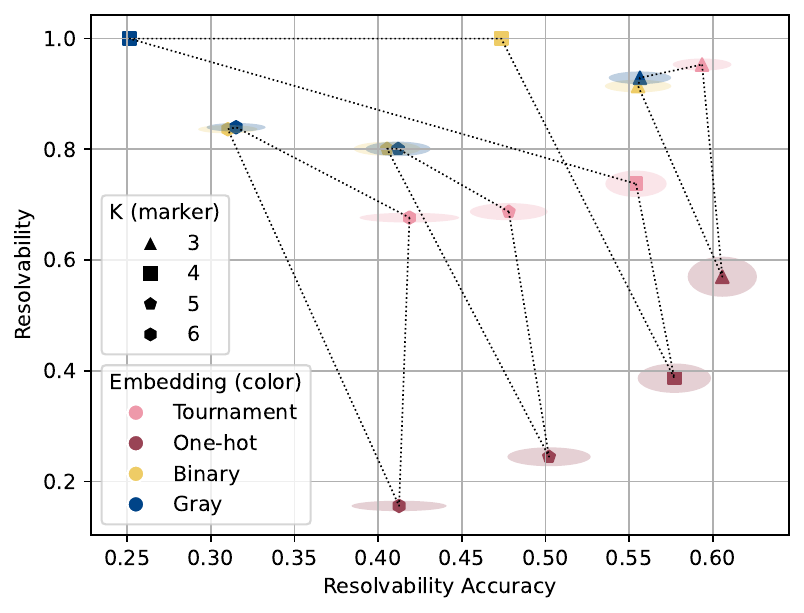}    
\end{minipage}
\begin{minipage}{.49\linewidth}
    \includegraphics[width=\linewidth,trim={1.5mm 1.5mm 1.5mm 1.5mm},clip]{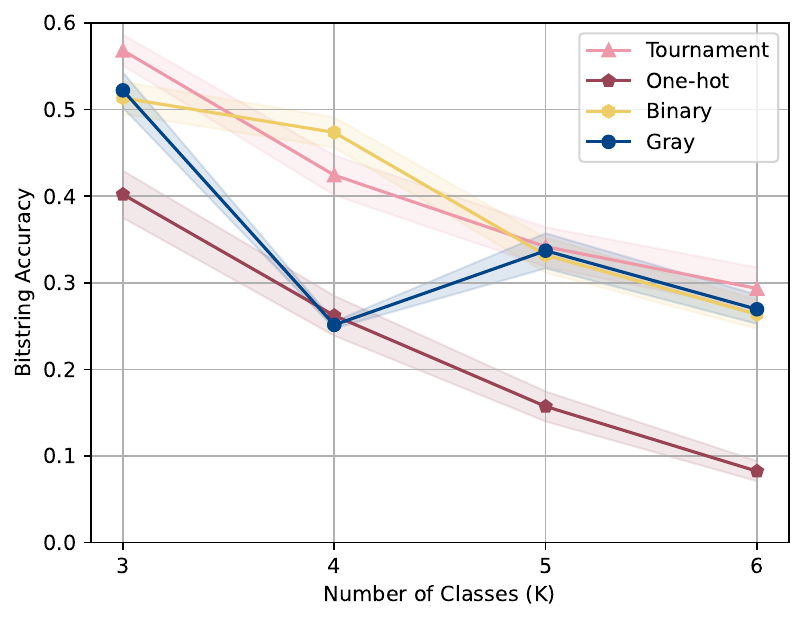}
\end{minipage}
\caption{
(\textit{left:}) Resolvability ratio $R$ versus shot-level resolvable accuracy $A_R^{\mathrm{shot}}$. Points show bootstrap means and confidence regions over matched experimental configurations. Tournament encoding occupies the high-$R$/high-$A_R^{\mathrm{shot}}$ region, while one-hot is primarily limited by resolvability and binary/Gray-code encodings by conditional shot accuracy.
(\textit{right:}) Raw bitstring shot accuracy $A_b^{\mathrm{shot}}$ as a function of class count $K$, with unresolvable predictions counted as incorrect. This fixed-shot metric summarizes the practical measurement-level reliability of each encoding.
}
\label{fig:metricsplots}

\end{figure*}

Table~\ref{tab:main_results} reports the main metrics aggregated by class count and encoding. The relationship between resolvability and shot-level resolvable accuracy is shown in the left panel of Fig.~\ref{fig:metricsplots}. Tournament encoding occupies the high-$R$/high-$A_R^{\mathrm{shot}}$ region across the tested class counts, indicating that it avoids both primary failure modes observed in the baselines. One-hot encoding maintains competitive conditional accuracy when a shot is resolvable, but its resolvability decreases rapidly with $K$. Binary and Gray-code encodings, by contrast, often maintain high resolvability, but their resolved shots are less accurate, particularly for larger class counts.

This separation is central to the proposed encoding. Resolvability alone is insufficient: binary and Gray-code measurements can frequently decode to valid labels while still producing weaker class decisions. 
Conditional accuracy alone is also insufficient: one-hot measurements can be accurate when valid, but many measurement shots are unresolvable. 
Tournament encoding combines redundant pairwise decisions with majority aggregation, yielding the strongest joint balance of validity and correctness under measurement among the tested encodings.

\subsection{Shot Quality Analysis}
\label{sec:shot_quality}
The right panel of Fig.~\ref{fig:metricsplots} shows how the valid-shot trade-off transfers to fixed-shot inference. The bitstring shot accuracy $A_b^{\mathrm{shot}}$ counts unresolvable outputs as incorrect, so it combines the two components shown in the left panel: how often an encoding produces a valid prediction and how often that valid prediction is correct. Tournament encoding maintains the strongest fixed-shot performance across the tested class counts, while one-hot accuracy degrades sharply as its resolvability decreases.

Binary and Gray-code encodings illustrate the opposite limitation. These encodings can achieve high resolvability, and binary encoding is fully resolvable at $K=4$, but their resolved-shot accuracy is substantially lower than that of tournament encoding at larger $K$. The resulting fixed-shot performance is therefore limited not by invalid outputs alone, but by the semantic fragility of compact bitstring decisions under sampling variability.

The top-1 simulation accuracy $T$ in Table~\ref{tab:main_results} further shows that continuous pre-measurement performance does not fully predict discrete inference quality. Encodings with similar top-1 simulation accuracy can differ substantially in $R$, $A_R^{\mathrm{shot}}$, and $A_b^{\mathrm{shot}}$, demonstrating that output encoding and measurement decoding are central to low-shot quantum classification.

\begin{figure*}[t]
    \centering
    \includegraphics[width=.95\linewidth]{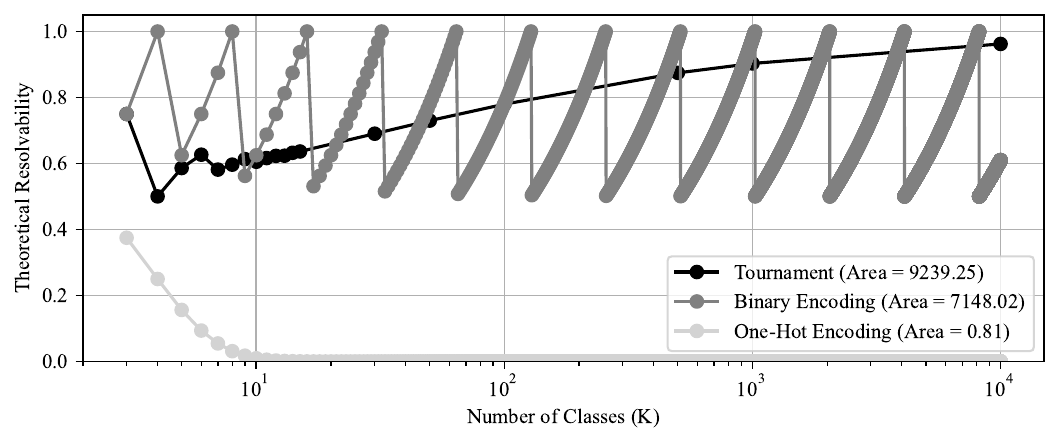}
    \vspace{-3mm}
    \caption{Figure comparing the $K\rightarrow\infty$ resolvability of one-hot encoding, both binary encodings, and our tournament encoding. Areas shown the legend account for total area under the resolvability curve, allowing for fair comparison between the oscillating binary curve and the smooth tournament and one-hot curves. Note that the plots start at $K=3$.}
    \label{fig:theoretic_shot_q}
\end{figure*}

\subsection{Scaling Behavior and Structural Limits}
\label{sec:validitycurves}

While our experiments focus on relatively small class counts ($K \leq 6$), the structural implications of each encoding become increasingly important as $K$ grows. 
Binary and Gray-code encodings exhibit a stepwise validity structure caused by the discrete jump in the number of output qubits, $\lceil \log_2 K\rceil$. When $K$ is not a power of two, the available bitstring space contains unused labels, so a fraction of measurement outcomes become invalid.
This leads to a bounded oscillatory degradation in resolvability, particularly when $\log_2 K$ is not an integer. 
For example, binary encoding achieves full validity at $K = 4$ (using 2 qubits), but drops to $62.5\%$ validity at $K = 5$ (using 3 qubits), as only 5 of the 8 possible bitstrings correspond to valid class labels. 
This produces oscillatory class-count dependence: binary and Gray-code encodings are fully valid when $K$ is a power of two, but lose validity immediately after each power-of-two transition.

This structural fragility implies that binary encodings are inherently sensitive to class count and qubit budget. For non-power-of-two $K$, the fraction of valid bitstrings decreases, and the probability of generating an unresolvable output rises sharply. 
This behavior is not merely empirical---it is a direct consequence of the encoding scheme's discrete nature. 

In Fig.~\ref{fig:theoretic_shot_q}, we present the oscillatory theoretical lower bounds for the resolvability probability of random bitstrings, together with the resolvability $K/2^K$ of one-hot encoding, which decreases exponentially with $K$, and finally the resolvability of our tournament encoding, which remains high and converges to unity as $K$ increases \cite{malinovsky2024roundrobintournamentsuniquemaximum}. 
The area under each resolvability curve summarizes how consistently an encoding maintains valid outputs across class counts; this structural AUC is highest for tournament encoding over the plotted range.

\section{Limitations}
\label{sec:lims}
Our proposed tournament-based encoding introduces a fundamental trade-off: quadratic qubit scaling with respect to the number of classes $K$. This requirement makes the approach impractical for large-scale problems until quantum hardware achieves significant improvements in qubit availability and fidelity. Consequently, all results in this paper are obtained under idealized conditions to isolate algorithmic behavior from hardware-specific noise. While this choice enables a clear evaluation of encoding strategies, robustness to real-device imperfections and resource constraints—both general and tournament-specific—remains an open challenge.

These constraints position our work as a theoretical analysis of output encodings rather than a direct path to near-term hardware deployment. The guarantees we provide, such as the convergence of resolvability to unity as $K \to \infty$ \citep{malinovsky2024roundrobintournamentsuniquemaximum}, are purely combinatorial and hold regardless of backend fidelity or noise. Our empirical evaluation under noiseless simulation demonstrates these properties in practice, but does not claim hardware readiness.

In addition, our experiments assume balanced datasets with clean labels. Class imbalance and semantic overlap introduce structural challenges: imbalance may bias majority voting toward dominant classes, while overlapping decision boundaries can increase the likelihood of cycles, which our current framework discards as ``unresolvable.'' These effects are not unique to quantum classifiers—they also affect classical one-vs-one schemes—but their impact on resolvability and accuracy under tournament aggregation remains an open question. We highlight these limitations explicitly and view extensions such as weighted voting, cycle-aware heuristics, and adaptive tie-breaking as promising directions for future work.

Future research should also explore strategies to mitigate quadratic scaling, such as hierarchical or sparse tournament structures, hybrid aggregation schemes, and alternative scoring mechanisms. Extensions inspired by classical tournament theory \cite{moon1968topics} (e.g., Condorcet-cycle handling, Schulze methods) offer promising directions for improving both efficiency and resolvability. Assuming continued progress in quantum hardware and deeper theoretical development, a large-scale experimental study on real quantum processors would be a natural next step. Such work is essential before deploying tournament-style models on high-dimensional datasets or production-level tasks.

\section{Conclusions}

In this paper, we take the first step toward improving the resolvability and accuracy of discrete outputs from multi-class PQC classifiers.  Achieving this goal has broader implications for quantum machine learning, as reducing sampling requirements removes a significant obstacle to quantum computing.
Our findings, supported by experiments, highlight a novel direction in quantum machine learning. We focus on designing models that yield resolvable and accurate discrete outputs more often by leveraging tournament solutions.

To achieve this, we propose a classical post-processing method for PQCs that maps the output space to a regular simplex, leading to the model learning a probabilistic directed graph over classes. Under deterministic inference, such models produce resolvable samples whenever there is a unique majority "winner," rather than only when an exact bitstring is produced. 

This effect upper bounds sampling needs as the number of classes increases while still producing highly accurate single-shot measurements, as compared to standard one-hot or bitstring based methods. 
Empirically, our results show our method achieves high resolvability and accuracy even with a single shot, outperforming other encodings in low-shot regimes while matching their performance under high sampling and simulation.

\section*{Acknowledgements}

This work was supported by the Swedish Research Council (VR) under grants 2023-05031 and 2022-06725. The funders played no role in study design, data collection, analysis and interpretation of data, or the writing of the manuscript. Computational and storage resources were provided by the National Academic Infrastructure for Supercomputing in Sweden (NAISS), funded by the Swedish Research Council, through resources hosted by the National Supercomputer Center at Linköping University, under project LiU-compute-2024-22.
Additional compute resources were provided by AISweden.

\section*{Author contributions}

A.D.H. conceived the tournament-based encoding framework, implemented the simulations, performed the experiments, analysed the results, prepared the figures and tables, and wrote the initial manuscript. J.~\AA.L. contributed to the theoretical framing, interpretation of the tournament-theoretic aspects, supervision, and manuscript revision. M.F. contributed to the machine-learning methodology, experimental design, supervision, interpretation of results, and manuscript revision. All authors discussed the results and approved the final manuscript.

\section*{Competing interests}

The authors declare no financial or non-financial competing interests.

\section*{Data availability}
\label{back:data_avail}

The datasets analysed in this study are publicly available. The MNIST handwritten digit dataset and Fashion-MNIST dataset were used for the classification experiments described in the Methods. The processed class subsets, random seeds, trained-model outputs, and aggregate data required to reproduce the figures and tables are available at \cite{github}.
Additional information required to reproduce the analyses is available from the corresponding author upon reasonable request.

\section*{Code availability}
\label{back:code_avail}

The custom code used to train the parameterized quantum circuits, implement the one-hot, binary, Gray-code, and tournament encodings, compute the shot-resolvability metrics, and generate the figures and tables is available at \cite{github}.
The repository includes source code, configuration files, random seeds, and instructions for reproducing the main numerical results.

\section*{Use of artificial intelligence tools}

Large language model tools were used for editorial assistance, including language refinement and consistency checks. The authors reviewed and approved all text, analyses, code, results, and scientific claims, and take full responsibility for the content of the manuscript.
\bibliography{references.bib}  
  
\appendix
\clearpage

\section{Full Testing Results}
\label{sec:fulldata}
In Table~\ref{tab:full_dataset}, we provide the same data provided in the main results but expanded by dataset. Similarly, in Tables
\ref{tab:full_circuit_1} and \ref{tab:full_circuit_2}, we provide the expansion over circuits. The remaining Tables \ref{tab:full_t}, \ref{tab:full_o}, \ref{tab:full_b}, and \ref{tab:full_g} are a robust expansion over circuits and datasets, meaning only the 5 seeds per run are used in the confidence intervals. 
The detailed results from all tests run up to this date, including individual seeded run results, can be found in the linked repository~\cite{github}.

\section{Ablations}\label{sec:ablations}

The main results focus on shot-level metrics because the primary goal of this work is to evaluate the quality of individual measured bitstrings under limited sampling. In the ablation studies, we report a broader diagnostic metric set that includes both shot-level and majority-vote accuracies. For any accuracy metric, the superscript ``shot'' denotes correctness at the level of individual measured bitstrings, while ``maj'' denotes correctness of the majority-vote prediction over a fixed set of measurement shots for each test example. Thus, $A_R^{\mathrm{shot}}$ and $A_b^{\mathrm{shot}}$ measure single-shot reliability, whereas $A_R^{\mathrm{maj}}$ and $A_b^{\mathrm{maj}}$ measure whether repeated samples recover the correct class after aggregation.

These majority-vote metrics were used as diagnostic quantities when selecting activation functions, optimization settings, and circuit-depth settings. They are included here to document the stability of these choices, but the main claims of the paper are based on the shot-level metrics reported in Section~\ref{sec:results}.
Majority-vote accuracies are omitted from the main results to keep the evaluation focused on single-shot inference, but are reported in the ablations because they are useful diagnostics for distinguishing unstable shot-level behaviour from settings that remain reliable after aggregation.

\subsection{Activation Functions}\label{sec:tempering}
\begin{table}[h!]
\scriptsize
    \centering
    \centering
    \caption{Averaged Friedman rank $F$ over all relevant statistics on CNN7 with $K=4$ using the edge method for minimum grad $g\in\{0.01,0.001\}$. Highest score was chosen as the optimal.}
    \vskip .15in
    \setlength\tabcolsep{3.2mm}
    \begin{tabular}{ccc}
        \toprule
        Function & $F|(g=0.01)$ & $F|(g=0.001)$ \\
        \midrule
        ERF & $\mathbf{5.0}$   & $4.375$ \\
        Linear & $2.875$   & $2.875$ \\
        Logistic & $4.725$   & $3.875$ \\
        Gudermannian & $4.25$   & $3.25$ \\
        \bottomrule
    \end{tabular}
    \label{tab:temp_ablate}
% \end{adjustbox}
\end{table}

\begin{figure}[b]
\centering
\includegraphics[width=\linewidth,trim={.60cm .65cm .6cm .60cm},clip]{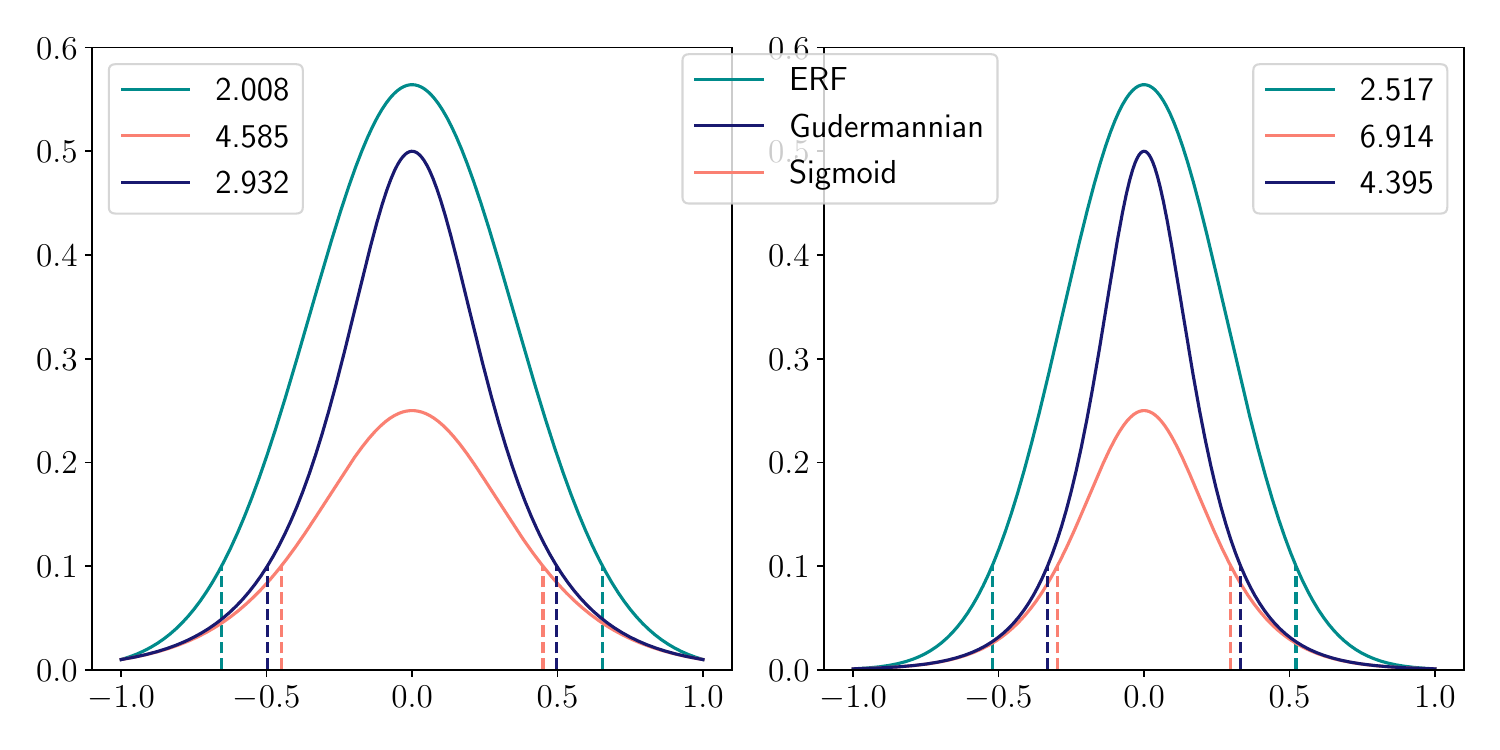}
\caption{
Derivatives of the candidate tempering functions after input rescaling. 
The left and right panels set the endpoint derivative at $\langle Z\rangle=\pm1$ to $0.01$ and $0.001$, respectively; legend values give the input scale factors required to achieve these minima. 
Dashed vertical lines mark where each derivative equals $0.1$, showing the width of the high-gradient region. 
With matched endpoint gradients, the logistic and Gudermannian functions have lower peak gradients and vanish sooner than the error function (ERF).}
\label{fig:activation}
\end{figure}

For all four encodings, the expectation values from the PQC are activated using a sigmoid function, inspired by the soft-threshold activation of equation (13) in Ref.~\cite{FELSBERG2009628}. 
This tempering reverses the monotonicity of the data and normalizes it, which both need to be done since the expectation value range for a quantum Pauli measurement is $[-1,1]$, and expectation values of $-1$ and $1$ are commonly used to represent a binary $1$ and $0$, respectively \citep{nielsen_quantum_2010, schuld_machine_2021}. Activating expectation values this way enables us to reason about them as the probabilities that their qubits, when discretized through measurement, will output $1$ as opposed to $0$.

The secondary goal in applying such a function is to ensure that the gradients returning to the circuit are minimal near expectation values of $-1$ and $1$, and maximal near $0$, since expectations near the extrema are more likely to discretize to either $1$ or $0$, respectively, and expectation values of $0$ operate like coin-flips when discretized.
Vanishing gradients from the sigmoid function have been a large enough problem in classical machine learning for them to be considered outdated \citep{ven2021regularizationreparameterizationavoidvanishing,roodschild2020new}, but in this use case, it provides exactly the behavior we want. Originally, the logistic function was chosen due to the ease of calculating its gradient \citep{goyal2020activation}, which, while efficient, may not lead to the optimal training behavior in quantum circuits.
\begin{table*}[ht]
\scriptsize
\centering
    \setlength\tabcolsep{2mm}
    \caption{Ablation on inference performance of noiseless-trained models using all six block circuit variants on simulated noisy hardware. Results are mean $\pm$ standard deviation over 5 seeds and 6 circuits.}
    \vskip .15in
\begin{tabular}{cccccccccc}
\toprule
 Dataset & $K$ & Method & ${A}_R^{\mathrm{shot}}$ & ${A}_R^{\mathrm{maj}}$ & $R$ & ${A}_b^{\mathrm{shot}}$ & ${A}_b^{\mathrm{maj}}$ & $T$ \\
\midrule
Digits & 3 & Tournament & $0.37 \pm 0.04$ & $0.43 \pm 0.11$ & $0.76 \pm 0.01$ & $0.28 \pm 0.03$ & $0.38 \pm 0.10$ & $0.69 \pm 0.16$ \\
Digits & 3 & One-hot & $0.33 \pm 0.00$ & $0.33 \pm 0.01$ & $0.38 \pm 0.03$ & $0.13 \pm 0.01$ & $0.00 \pm 0.00$ & $0.67 \pm 0.20$ \\
Digits & 4 & Tournament & $0.28 \pm 0.02$ & $0.34 \pm 0.05$ & $0.53 \pm 0.02$ & $0.15 \pm 0.02$ & $0.03 \pm 0.06$ & $0.76 \pm 0.25$ \\
Digits & 4 & One-hot & $0.25 \pm 0.00$ & $0.25 \pm 0.01$ & $0.27 \pm 0.03$ & $0.08 \pm 0.01$ & $0.00 \pm 0.00$ & $0.76 \pm 0.26$ \\
Fashion & 3 & Tournament & $0.39 \pm 0.05$ & $0.50 \pm 0.11$ & $0.77 \pm 0.02$ & $0.30 \pm 0.04$ & $0.46 \pm 0.11$ & $0.70 \pm 0.16$ \\
Fashion & 3 & One-hot & $0.33 \pm 0.00$ & $0.33 \pm 0.01$ & $0.37 \pm 0.05$ & $0.13 \pm 0.02$ & $0.00 \pm 0.00$ & $0.69 \pm 0.20$ \\
Fashion & 4 & Tournament & $0.26 \pm 0.02$ & $0.29 \pm 0.05$ & $0.52 \pm 0.02$ & $0.14 \pm 0.01$ & $0.02 \pm 0.04$ & $0.71 \pm 0.24$ \\
Fashion & 4 & One-hot & $0.25 \pm 0.00$ & $0.25 \pm 0.01$ & $0.30 \pm 0.02$ & $0.08 \pm 0.01$ & $0.00 \pm 0.00$ & $0.73 \pm 0.21$ \\
\bottomrule

\end{tabular}\medskip
    \label{tab:noise}
\end{table*}

There are many functions which have the required shape, with the biggest difference between them being their domains relative to their asymptotes as none reach diminished gradients in the domain $[-1,1]$.
Because of this, the inputs to the functions need to be scaled to make full use of this vanishing effect.
This scaling can be such that the minimum gradient returning to the circuit is arbitrarily close to $0$, but the more this scaling is applied, the more of the input domain receives very little gradient, as shown in Fig.~\ref{fig:activation}.
In this study, we ablated over three sigmoid like functions - namely, the logistic function, the error function, and the Gudermannian function \citep{gambini2024structural} - and two minimum gradient levels for each - namely, $0.01$ and $0.001$.
To calculate the scaling, we simply find the input value to the first derivative of each function that gives the minimum value we set.

To determine which sigmoid-like function to use for the main results, we performed our ablation process on the CNN7 block \cite{sim_expressibility_2019} using $K=4$ on the MNIST Digits dataset, shown in Table~\ref{tab:temp_ablate}. We compared ERF, Gudermannian, and the logistic function at minimal gradient values of~ $0.01$~and~
$0.001$, as well as a linear monotonicity-reversing normalization.

The scores presented are Friedman-scores computed over a range metrics: both sets of shot and majority-vote accuracies, the threshold accuracy $T$ and average distance between the top-two predictions, as well as the resolvability $R$. We use the Friedman-scores to decide on the best tempering without focusing on a single metric.

\subsection{Hardware Noise}\label{sec:noise}

Because the compared encodings are evaluated on matched PQC architectures and initializations, the main experiments are designed to isolate differences due to decoding and post-processing rather than circuit topology. Hardware noise can nevertheless affect the learned and measured output distributions, so we include noise-model experiments as a supplementary robustness check.
To assess how much noise affects our relative performance, we ran our inference suite on both datasets with all of the circuit blocks with $K\in\{3,4\}$ using noise models provided by IBM QisKit \citep{Qiskit2024}. This inference was performed using the same learned parameters trained under noiseless-simulation to produce the results provided in Section~\ref{sec:results}. 

These noise models allow PennyLane to simulate the noise of real-world IBM hardware. Given the differing sizes of the circuits, we use the noise models of different retired IBM machines for different values of $K$—namely IBM Belem Version 2 for $K=3$ and IBM Oslo for $K=4$. The varying levels of noise for the varying hardware make comparing performance across different values of $K$ less productive, however the results in Table~\ref{tab:noise} still allow for a fair comparison of the primary results between the tournament and one-hot post-processing methods. 

It can be clearly seen that hardware noise has a universally negative effect on the performance of even pretrained PQCs, however, the effectiveness of the tournament method over the one-hot method is still clearly visible. The relative performance between the two methods remains either identical or even improves for the tournament method. This is especially evident for the resolvable-only majority-vote accuracy $R_a$, which remains better than guessing under the tournament method, but hovers near guessing level for the one-hot method. This improvement is attained using less samples—as evidenced by the superior resolvability $R$. 
These supplementary experiments indicate that the tournament method preserves its relative advantage over one-hot decoding in the tested noise models. We interpret these results as evidence of robustness within the simulated conditions considered here, rather than as a claim of hardware readiness.
Updating to current backends would not affect the tournament-theoretic motivation of the decoding rule, but may affect the empirical noise-model values.

\subsection{Circuit Depth}\label{sec:depth}
In order to test the assumption that the results of our post-processing method are unrelated to circuit depth, we ran an ablation over the number of circuit block layers. We focus on $K\in\{4,5\}$ for the circuits CNN8, SU(4), and SEL-X, and test the same metrics with layers $L\in\{1,3,6\}$. The results provided in Table~\ref{tab:depth} show that, indeed, the discrete outputs of circuits trained with our tournament method remain more resolvable and similarly accurate even as circuit depth varies with unsurprising minor changes in overall accuracy between both methods.
\begin{table*}[th]
\scriptsize
\centering
    \setlength\tabcolsep{3mm}
    \caption{Ablation showing the differing performance of circuits trained and tested with a number of 2-ring layers $L\in\{1,3,4,6\}$. The main results of the paper are attained using $L=4$. Blocks SU(4), CNN8, and SEL-X, the MNIST Digits dataset, and $K\in\{4,5\}$ were used for the ablation.}
    \vskip .15in
\begin{tabular}{ccccccccccc}
\toprule
 $K$ & Block & $L$ & Method & ${A}_R^{\mathrm{shot}}$ & ${A}_R^{\mathrm{maj}}$ & $R$ & ${A}_b^{\mathrm{shot}}$ & ${A}_b^{\mathrm{maj}}$ & $T$ \\

\midrule
4 &  CNN8 & 1 &     Tournament & $58.26$ & $82.68$ & $0.62$ & $38.52$ & $53.44$ & $86.31$ \\
4 &  CNN8 & 1 &   One-hot & $56.04$ & $80.06$ & $0.39$ & $23.91$ &  $6.73$ & $83.42$ \\
4 & SU(4) & 1 &     Tournament & $59.86$ & $84.71$ & $0.65$ & $40.73$ & $61.31$ & $87.15$ \\
4 & SU(4) & 1 &   One-hot & $57.64$ & $83.38$ & $0.42$ & $26.44$ & $10.88$ & $85.75$ \\
4 & SEL-X & 1 &     Tournament & $32.45$ & $38.25$ & $0.54$ & $19.46$ & $16.58$ & $47.49$ \\
4 & SEL-X & 1 &   One-hot & $37.63$ & $42.32$ & $0.27$ &  $13.6$ &  $3.73$ & $50.33$ \\
\midrule
4 &  CNN8 & 3 &     Tournament & $61.53$ & $87.62$ & $0.70$ & $44.78$ & $70.92$ & $89.91$ \\
4 &  CNN8 & 3 &   One-hot & $62.57$ & $88.97$ & $0.50$ & $32.76$ & $22.59$ &  $90.4$ \\
4 & SU(4) & 3 &     Tournament & $63.43$ & $89.26$ & $0.75$ & $48.52$ & $81.78$ & $90.65$ \\
4 & SU(4) & 3 &   One-hot & $64.96$ & $89.04$ & $0.54$ & $36.67$ & $36.68$ & $90.82$ \\
4 & SEL-X & 3 &     Tournament & $44.28$ & $70.79$ & $0.53$ &  $24.5$ & $12.45$ & $76.49$ \\
4 & SEL-X & 3 &   One-hot & $39.15$ & $64.28$ & $0.30$ & $12.46$ &  $0.01$ & $76.67$ \\
\midrule
4 & CNN8  & 4 &     Tournament & $64.03$ & $89.17$ & $0.73$ & $48.2$ & $80.43$ & $90.54$ \\
4 & CNN8  & 4 &   One-hot & $64.38$ & $89.47$ & $0.54$ & $36.35$ & $35.86$ & $90.9$ \\
4 & SU(4) & 4 &     Tournament & $63.51$ & $89.21$ & $0.74$ & $48.38$ & $81.04$ & $90.91$ \\
4 & SU(4) & 4 &   One-hot & $64.55$ & $89.86$ & $0.57$ & $38.07$ & $44.31$ & $91.48$ \\
4 & SEL-X & 4 &     Tournament & $46.86$ & $74.9$  & $0.55$ & $26.94$ & $17.6$ & $81.25$ \\
4 & SEL-X & 4 &   One-hot & $48.04$ & $75.95$ & $0.37$ & $19.32$ & $2.59$ &  $84.21$ \\
\midrule
4 &  CNN8 & 6 &     Tournament & $62.27$ & $89.57$ & $0.75$ & $47.71$ & $81.91$ & $91.15$ \\
4 &  CNN8 & 6 &   One-hot & $62.14$ & $89.62$ & $0.53$ & $34.18$ & $28.36$ & $91.34$ \\
4 & SU(4) & 6 &     Tournament & $61.03$ & $89.64$ & $0.75$ & $46.37$ & $82.74$ & $91.02$ \\
4 & SU(4) & 6 &   One-hot & $62.12$ & $89.72$ & $0.53$ & $34.12$ & $28.04$ & $91.57$ \\
4 & SEL-X & 6 &     Tournament & $48.85$ & $81.44$ & $0.59$ & $29.51$ & $26.75$ & $87.13$ \\
4 & SEL-X & 6 &   One-hot & $50.42$ & $82.67$ & $0.40$ & $21.02$ &  $1.48$ & $87.96$ \\
\midrule
5 &  CNN8 & 1 &     Tournament & $47.67$ & $71.91$ & $0.62$ & $30.72$ & $34.38$ &  $75.6$ \\
5 &  CNN8 & 1 &   One-hot & $41.57$ & $63.17$ & $0.33$ & $14.78$ &  $0.21$ & $69.79$ \\
5 & SU(4) & 1 &     Tournament & $48.47$ & $74.03$ & $0.63$ & $31.46$ & $38.67$ &  $77.0$ \\
5 & SU(4) & 1 &   One-hot & $43.64$ & $66.19$ & $0.35$ & $16.44$ &  $0.41$ & $71.76$ \\
5 & SEL-X & 1 &     Tournament &  $26.5$ & $32.99$ & $0.58$ & $15.88$ &  $7.87$ & $36.28$ \\
5 & SEL-X & 1 &   One-hot & $27.91$ & $30.81$ & $0.21$ &   $7.5$ &   $0.5$ & $32.07$ \\
\midrule
5 &  CNN8 & 3 &     Tournament & $52.19$ &  $81.1$ & $0.65$ & $34.98$ & $50.46$ & $83.39$ \\
5 &  CNN8 & 3 &   One-hot & $51.93$ & $81.48$ & $0.37$ & $20.46$ &  $0.94$ & $84.81$ \\
5 & SU(4) & 3 &     Tournament & $54.35$ & $82.01$ & $0.67$ & $37.52$ & $58.74$ & $84.78$ \\
5 & SU(4) & 3 &   One-hot & $56.11$ & $84.36$ & $0.40$ & $23.76$ &   $4.0$ & $86.78$ \\
5 & SEL-X & 3 &     Tournament &  $25.7$ & $41.61$ & $0.58$ & $15.11$ &  $1.38$ & $47.94$ \\
5 & SEL-X & 3 &   One-hot & $25.22$ & $39.07$ & $0.21$ &  $5.55$ &   $0.0$ & $48.38$ \\
\midrule
5 & CNN8  & 4 &     Tournament & $53.71$ & $80.97$ & $0.67$ & $37.2$  & $57.27$ & $83.77$ \\
5 & CNN8  & 4 &   One-hot & $53.68$ & $83.28$ & $0.39$ & $22.16$ &  $1.97$ & $85.33$ \\
5 & SU(4) & 4 &     Tournament & $55.75$ & $83.55$ & $0.67$ & $38.39$ & $57.95$ & $85.05$ \\
5 & SU(4) & 4 &   One-hot & $56.35$ & $85.33$ & $0.40$ & $24.13$ &  $4.31$ & $87.78$ \\
5 & SEL-X & 4 &     Tournament & $29.23$ & $49.57$ & $0.58$ & $17.21$ &  $2.25$ & $55.47$ \\
5 & SEL-X & 4 &   One-hot & $27.61$ & $45.6$  & $0.22$ & 6 $.55$ &   $0.0$ & $55.76$ \\
\midrule
5 &  CNN8 & 6 &     Tournament & $53.99$ & $82.15$ & $0.67$ & $37.39$ & $58.81$ & $84.53$ \\
5 &  CNN8 & 6 &   One-hot & $55.09$ & $84.74$ & $0.39$ &  $23.0$ &   $2.4$ & $87.23$ \\
5 & SU(4) & 6 &     Tournament & $56.66$ & $84.74$ & $0.70$ & $40.48$ & $66.49$ & $86.01$ \\
5 & SU(4) & 6 &   One-hot & $57.28$ & $86.03$ & $0.41$ & $25.42$ &  $5.93$ & $88.14$ \\
5 & SEL-X & 6 &     Tournament & $30.71$ & $47.26$ & $0.58$ & $18.29$ &   $7.3$ & $57.62$ \\
5 & SEL-X & 6 &   One-hot & $32.46$ & $56.96$ & $0.26$ &  $8.85$ &   $0.0$ & $68.98$ \\

\bottomrule

\end{tabular}\medskip
    \label{tab:depth}
\end{table*}

The larger relative improvement observed for shallower circuits suggests that tournament decoding may be useful in low-depth regimes, where hardware noise places practical limits on circuit depth. This observation is suggestive rather than conclusive, since the depth ablation itself is performed in simulation.

\subsection{Optimization}\label{sec:optablate}
To ascertain the best optimization strategy before running the full experimental suite, we ran an ablation across two optimizers, four learning rate schedulers, and three learning rates. The two optimizers are standard stochastic gradient descent (SGD) \cite{robbins1951stochastic} and Adam \cite{kingma2017adammethodstochasticoptimization}.
The learning rate schedulers we tested were an exponential scheduler \cite{li2019exponential}, a cosine scheduler \cite{loshchilov2016sgdr}, a piecewise scheduler \cite{goyal2017accurate}, and no (or constant) scheduler.

For the exponential scheduler, there were ten total transition steps over the full six epochs, with a decay rate of $0.9$. For the cosine scheduler, the number of steps was simply the number of training steps. For the piecewise scheduler, there were three transition steps with scale factors of $0.1$ and $0.01$.

We first ran all the tests on $K=3$ with the tournament method on the CNN7 block, shown in the top of Table~\ref{tab:opt}. To average over all the metrics, we look at the Friedman Rank (F-Rank) of each optimization strategy, which ranks the columns and averages the ranks over the rows \citep{friedman1937ranks}. Due to the tie between the piecewise scheduler and constant scheduler with the Adam optimizer, we opted to run a second set on the Adam optimizer with $K=5$ instead. This test is shown in the bottom of Table~\ref{tab:opt}. As exponential decay with a learning rate of $0.01$ ranked best for $K=5$, and nearly as well as piecewise and constant for $K=3$, this was chosen as the optimal setup.
\begin{table*}[t]
\scriptsize
\centering
    \setlength\tabcolsep{3mm}
    \caption{Ablation with $K \in \{3,5\}$ over optimizers (Opt), learning rate schedulers (LRS), and learning rates (LR). Schedulers used include Exponential Decay \emph{exp}, Cosine Decay \emph{cos}, Piecewise Constant \emph{step}, Constant \emph{reg}, and Linear Decay \emph{lin}. }
    \vskip .15in
\begin{tabular}{cccccccccccc}
\toprule
% K & Opt & LRS & LR & F-Rank  & \multicolumn{3}{c}{Valid} & \multicolumn{2}{c}{Raw Bitstring} & \multicolumn{1}{c}{Simulation} \\
%  \cmidrule(rl){5-7} \cmidrule(rl){8-9}\cmidrule(rl){10-10}
%   & &  & $(\uparrow)$ & ${A}_R^{\mathrm{shot}}$ & ${A}_R^{\mathrm{maj}}$ & $R$ & ${A}_b^{\mathrm{shot}}$ & ${A}_b^{\mathrm{maj}}$ & $T$ \\
K & Opt & LRS & LR & F-Rank  & $(\uparrow)$ & ${A}_R^{\mathrm{shot}}$ & ${A}_R^{\mathrm{maj}}$ & $R$ & ${A}_b^{\mathrm{shot}}$ & ${A}_b^{\mathrm{maj}}$ & $T$ \\
\midrule
3 & SGD & \emph{exp} & 0.01 &    $14.875$ & $54.15$ & $73.73$ & $116.36$ & $47.02$ & $72.7$ & $78.33$ \\
3 & SGD & \emph{exp} & 0.001 &   $5.375$ & $50.88$ & $68.84$ & $125.53$ & $41.6$ & $63.08$ & $75.02$ \\
3 & SGD & \emph{exp} & 0.0001 &  $2.0$ & $42.04$ & $50.62$ & $129.33$ & $33.5$ & $44.78$ & $57.68$ \\
3 & SGD & \emph{reg} & 0.01 &    $17.0$ & $54.16$ & $73.77$ & $116.32$ & $47.04$ & $72.77$ & $78.39$ \\
3 & SGD & \emph{reg} & 0.001 &   $7.625$ & $50.92$ & $68.99$ & $125.43$ & $41.66$ & $63.27$ & $74.84$ \\
3 & SGD & \emph{reg} & 0.0001 &  $3.125$ & $42.1$ & $50.76$ & $129.37$ & $33.53$ & $44.85$ & $58.29$ \\
3 & SGD & \emph{step} & 0.01 &   $17.0$ & $54.16$ & $73.77$ & $116.32$ & $47.04$ & $72.77$ & $78.39$ \\
3 & SGD & \emph{step} & 0.001 &  $7.625$ & $50.92$ & $68.99$ & $125.43$ & $41.66$ & $63.27$ & $74.84$ \\
3 & SGD & \emph{step} & 0.0001 & $3.125$ & $42.1$ & $50.76$ & $129.37$ & $33.53$ & $44.85$ & $58.29$ \\
3 & SGD & \emph{cos} & 0.01 &    $16.5$ & $54.16$ & $73.77$ & $116.32$ & $47.04$ & $72.77$ & $\textbf{78.4}$ \\
3 & SGD & \emph{cos} & 0.001 &   $6.875$ & $50.92$ & $68.98$ & $125.43$ & $41.66$ & $63.23$ & $74.85$ \\
3 & SGD & \emph{cos} & 0.0001 &  $2.375$ & $42.09$ & $50.74$ & $129.36$ & $33.53$ & $44.84$ & $58.28$ \\
3 & SGD & \emph{lin} & 0.01 &    $15.625$ & $54.15$ & $73.75$ & $116.35$ & $47.02$ & $72.71$ & $78.35$ \\
3 & SGD & \emph{lin} & 0.001 &   $6.125$ & $50.89$ & $68.84$ & $125.52$ & $41.61$ & $63.11$ & $75.03$ \\
3 & SGD & \emph{lin} & 0.0001 &  $2.5$ & $42.05$ & $50.65$ & $129.34$ & $33.5$ & $44.79$ & $57.72$ \\
3 & Adam & \emph{exp} & 0.01 &   $16.75$ & $54.62$ & $69.29$ & $\textbf{114.15}$ & $48.49$ & $68.54$ & $76.66$ \\
3 & Adam & \emph{exp} & 0.001 &  $19.875$ & $\textbf{55.49}$ & $\textbf{75.16}$ & $114.41$ & $48.92$ & $74.25$ & $78.26$ \\
3 & Adam & \emph{exp} & 0.0001 & $9.5$ & $51.7$ & $70.33$ & $123.64$ & $42.71$ & $65.42$ & $75.14$ \\
3 & Adam & \emph{reg} & 0.01 &   $17.5$ & $54.65$ & $69.3$ & $114.19$ & $48.49$ & $68.63$ & $76.64$ \\
3 & Adam & \emph{reg} & 0.001 &  $\textbf{21.125}$ & $\textbf{55.49}$ & $75.07$ & $114.35$ & $\textbf{48.94}$ & $\textbf{74.29}$ & $78.27$ \\
3 & Adam & \emph{reg} & 0.0001 & $12.125$ & $51.76$ & $70.52$ & $123.53$ & $42.78$ & $65.6$ & $75.17$ \\
3 & Adam & \emph{step} & 0.01 &  $17.5$ & $54.65$ & $69.3$ & $114.19$ & $48.49$ & $68.63$ & $76.64$ \\
3 & Adam & \emph{step} & 0.001 & $\textbf{21.125}$ & $\textbf{55.49}$ & $75.07$ & $114.35$ & $\textbf{48.94}$ & $\textbf{74.29}$ & $78.27$ \\
3 & Adam & \emph{step} & 0.0001 &$12.125$ & $51.76$ & $70.52$ & $123.53$ & $42.78$ & $65.6$ & $75.17$ \\
3 & Adam & \emph{cos} & 0.01 &   $17.375$ & $54.65$ & $69.29$ & $114.19$ & $48.49$ & $68.64$ & $76.65$ \\
3 & Adam & \emph{cos} & 0.001 &  $21.0$ & $\textbf{55.49}$ & $75.08$ & $114.35$ & $\textbf{48.94}$ & $\textbf{74.29}$ & $78.27$ \\
3 & Adam & \emph{cos} & 0.0001 & $11.625$ & $51.76$ & $70.52$ & $123.53$ & $42.78$ & $65.61$ & $75.16$ \\
3 & Adam & \emph{lin} & 0.01 &   $17.75$ & $54.63$ & $69.3$ & $\textbf{114.15}$ & $48.5$ & $68.59$ & $76.66$ \\
3 & Adam & \emph{lin} & 0.001 &  $20.5$ & $\textbf{55.49}$ & $\textbf{75.16}$ & $114.41$ & $48.93$ & $74.27$ & $78.26$ \\
3 & Adam & \emph{lin} & 0.0001 & $10.375$ & $51.71$ & $70.37$ & $123.63$ & $42.72$ & $65.45$ & $75.14$ \\
\midrule
5 & Adam & \emph{exp} & 0.01   & $\textbf{11.0}$ & $\textbf{49.25}$ & $80.43$ & $151.77$ & $\textbf{33.16}$ & $47.5$ & $\textbf{83.99}$ \\
5 & Adam & \emph{exp} & 0.001  & $5.125$ & $47.62$ & $78.95$ & $155.31$ & $31.41$ & $40.19$ & $82.9$ \\
5 & Adam & \emph{exp} & 0.0001   & $2.375$ & $36.63$ & $65.06$ & $166.9$ & $22.41$ & $13.36$ & $72.54$ \\
5 & Adam & \emph{cos} & 0.01    & $9.625$ & $49.23$ & $80.36$ & $151.79$ & $33.14$ & $47.49$ & $83.92$ \\
5 & Adam & \emph{cos} & 0.001  & $6.875$ & $47.62$ & $78.92$ & $155.21$ & $31.43$ & $40.14$ & $82.99$ \\
5 & Adam & \emph{cos} & 0.0001   & $3.75$ & $36.75$ & $64.85$ & $166.84$ & $22.48$ & $13.67$ & $72.7$ \\
5 & Adam & \emph{step} & 0.01   & $9.625$ & $49.23$ & $80.36$ & $151.79$ & $33.14$ & $47.49$ & $83.92$ \\
5 & Adam & \emph{step} & 0.001 & $6.875$ & $47.62$ & $78.92$ & $155.21$ & $31.43$ & $40.14$ & $82.99$ \\
5 & Adam & \emph{step} & 0.0001  & $3.75$ & $36.75$ & $64.85$ & $166.84$ & $22.48$ & $13.67$ & $72.7$ \\
5 & Adam & \emph{reg} & 0.01    & $10.625$ & $49.24$ & $\textbf{80.44}$ & $151.78$ & $33.15$ & $\textbf{47.54}$ & $83.94$ \\
5 & Adam & \emph{reg} & 0.001  & $6.75$ & $47.62$ & $78.96$ & $155.22$ & $31.42$ & $40.33$ & $82.98$ \\
5 & Adam & \emph{reg} & 0.0001   & $3.5$ & $36.75$ & $65.11$ & $166.87$ & $22.48$ & $13.61$ & $72.71$ \\
5 & Adam & \emph{lin} & 0.01  & $10.25$ & $49.23$ & $80.36$ & $\textbf{151.76}$ & $\textbf{33.16}$ & $47.46$ & $83.97$ \\
5 & Adam & \emph{lin} & 0.001  & $5.25$ & $47.6$ & $78.78$ & $155.27$ & $31.41$ & $40.32$ & $82.9$ \\
5 & Adam & \emph{lin} & 0.0001  & $2.875$ & $36.65$ & $64.97$ & $166.9$ & $22.42$ & $13.44$ & $72.57$ \\
\bottomrule
\end{tabular}\medskip
    \label{tab:opt}
\end{table*}

\section{QML Block Descriptions}
\label{sec:blocks}
Here we present information about the blocks used in the 2-qubit ring structure. In this section, we will simply summarize the findings of the introductory works to justify their usage in this paper.

The first four of the blocks were found in Ref.~\cite{hur_quantum_2022} which also showed promising results in  performed experiments.
In that paper, the reason each block was chosen was explained succinctly.
The CNN7 and CNN8 blocks were first introduced as 4-qubit error-correcting encoders \cite{johnson2017qvectoralgorithmdevicetailoredquantum}.
They showed better expressibility \cite{sim_expressibility_2019}, leading to them being chosen in Ref.~\cite{hur_quantum_2022}.
Expressibility, in the context of QML, is a measure of the ability of a circuit to produce a wide range of quantum states.

The SO(4) block can implement an arbitrary SO(4) operation \cite{wei_decomposition_2012}, and can be used to construct a fully entangled VQE. 
Similarly, the SU(4) block can implement any arbitrary 2-qubit rotation  \cite{vatan2004optimal, maccormack2020branchingquantumconvolutionalneural}.

Strongly Entangling Layers (SEL) is a popular multi qubit gate-operation that is available as a callable function in the popular quantum computing package PennyLane \cite{bergholm2022pennylaneautomaticdifferentiationhybrid}. The setup was invented in a paper by several of the authors responsible for the creation of PennyLane \cite{schuld_circuit-centric_2020}, and has seen much use due to simplicity and expressibility.

Note that for the Strongly Entangling Layers block all single-qubit operations are applied before the ring of two-qubit operations rather than in alternating full block rings like in CNN7, CNN8, SO(4) and SU(4), as visualized in Fig.~\ref{fig:training}. We included this block in our analysis so as to demonstrate the efficacy of the tournament encoding independent of the 2-qubit ring structure.
For even greater fairness, we include two versions using the two most common parameter-free 2-qubit operations, namely the CNOT gate and CZ gate. 

\begin{figure}[bt]
    \centering
        \begin{minipage}{\linewidth}
            \includegraphics[width=.9\linewidth,trim={3.7cm 1.5cm 3.8cm 1.5cm},clip]{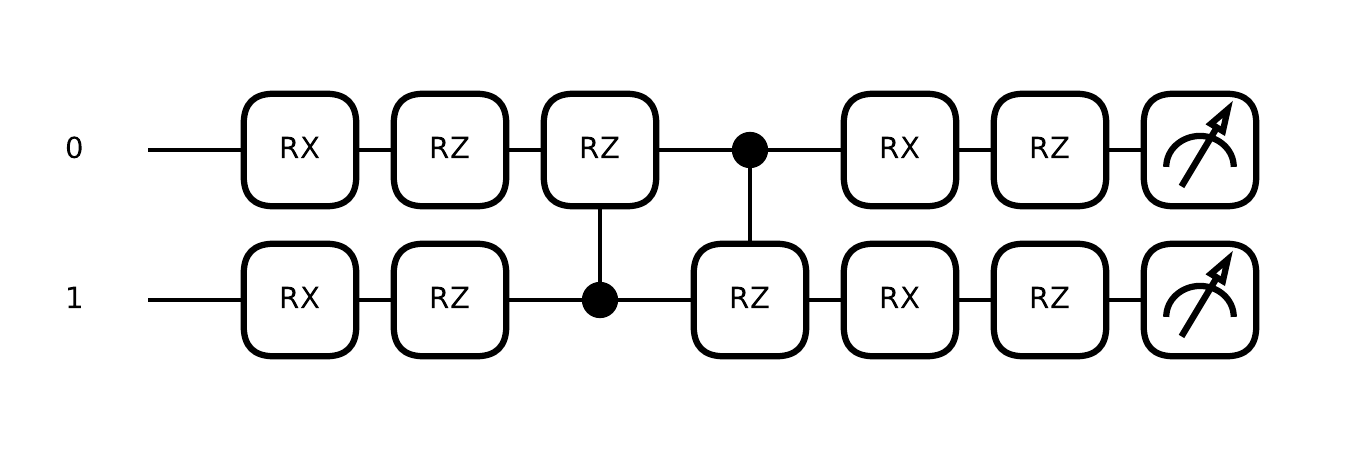}
            \caption{CNN7 Block \cite{sim_expressibility_2019}, \\ as modified by Ref.~\cite{hur_quantum_2022}.}
            \label{fig:cnn7}
        \end{minipage}\bigskip

        \begin{minipage}{\linewidth}
            \includegraphics[width=.9\linewidth,trim={3.7cm 1.5cm 3.8cm 1.5cm},clip]{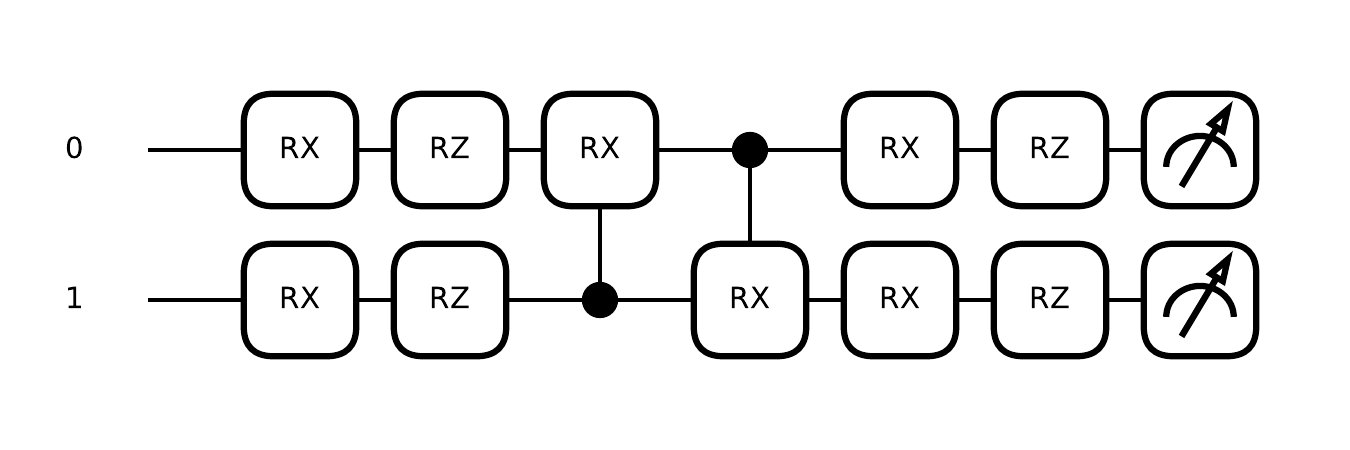}
            \caption{CNN8 Block \cite{sim_expressibility_2019}, \\ as modified by Ref.~\cite{hur_quantum_2022}.}
            \label{fig:cnn8}
        \end{minipage}\bigskip

        \begin{minipage}{\linewidth}
            \includegraphics[width=.85\linewidth,trim={3.7cm 1.5cm 3.7cm 1.5cm},clip]{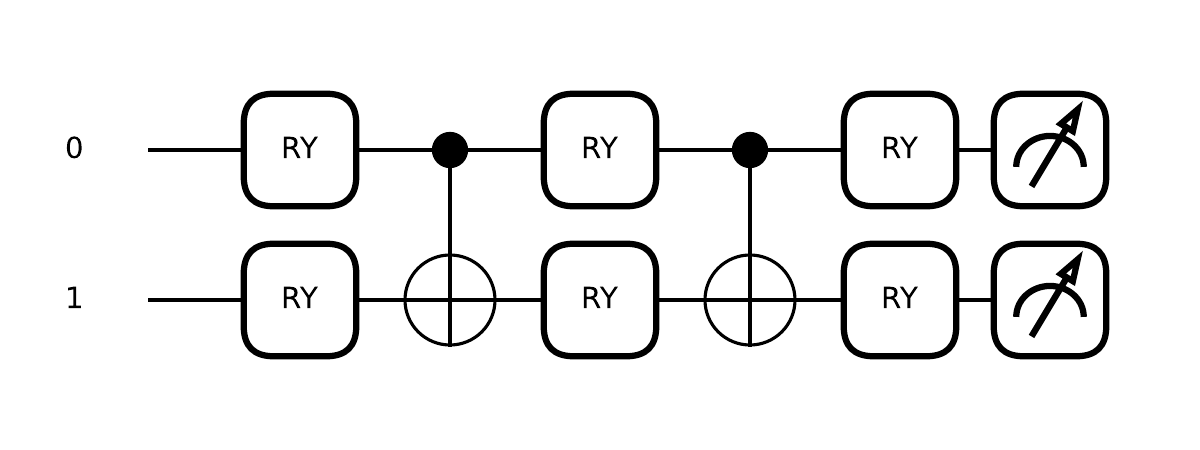}
            \caption{SO(4) Block \cite{wei_decomposition_2012}, \\ as modified by Ref.~\cite{hur_quantum_2022}.}
            \label{fig:uso4}
        \end{minipage}\bigskip

        \begin{minipage}{\linewidth}
            \includegraphics[width=\linewidth,trim={3.7cm 1.5cm 3.7cm 1.5cm},clip]{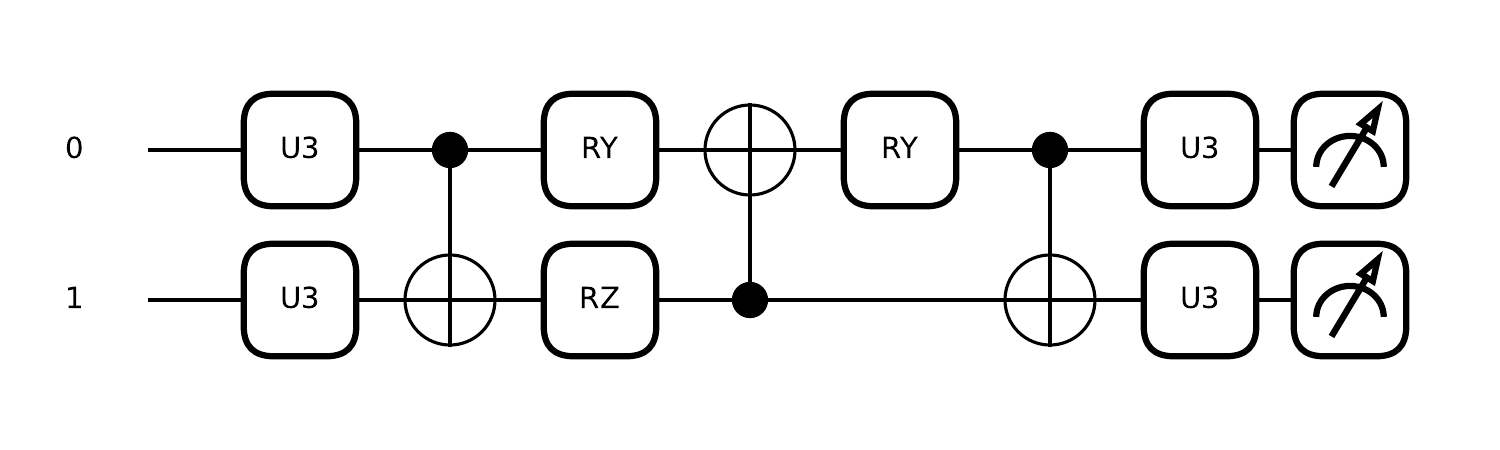}
            \caption{SU(4) Block \cite{vatan2004optimal}, \\ as modified by Ref.~\cite{hur_quantum_2022}.}
            \label{fig:usu4}
        \end{minipage}\bigskip
        
        \begin{minipage}{\linewidth}
            \includegraphics[width=.85\linewidth,trim={3.7cm 1.5cm 3.9cm 1.5cm},clip]{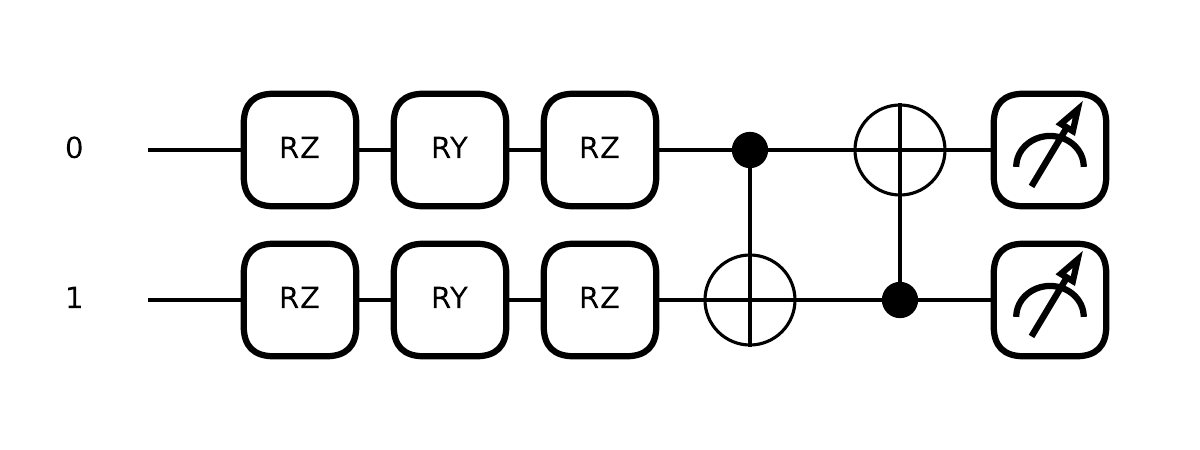}
            \caption{SEL Block with \\CNOT imprimitive (Sel-X) \cite{schuld_circuit-centric_2020}.}
            \label{fig:selx}
        \end{minipage}\bigskip
        
        \begin{minipage}{\linewidth}
            \includegraphics[width=.85\linewidth,trim={3.7cm 1.5cm 3.9cm 1.5cm},clip]{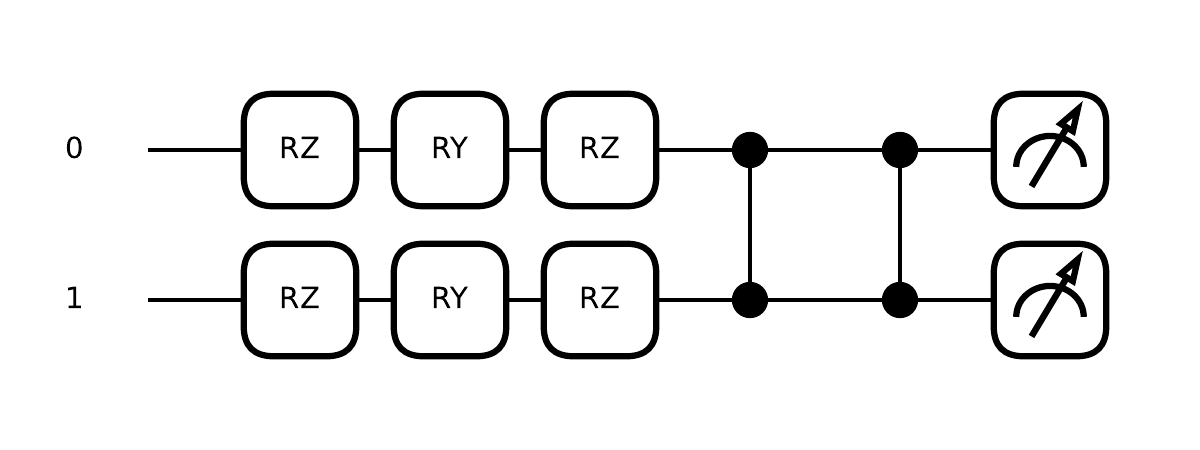}
            \caption{SEL Block with \\CZ imprimitive (Sel-Z) \cite{schuld_circuit-centric_2020}.}
            \label{fig:selz}
        \end{minipage}\bigskip
    \label{fig:circdiagrams}
\end{figure}

\begin{table*}[t]
\scriptsize
    \centering
    \caption{Expanded metrics aggregated by class count, encoding, and dataset. The table reports resolvability ratio $R$, shot-level resolvable accuracy $A_R^{\mathrm{shot}}$, raw bitstring shot accuracy $A_b^{\mathrm{shot}}$, and continuous top-1 simulation accuracy $T$. The first three metrics evaluate discrete measurement outcomes: $R$ measures how often a shot yields a valid prediction, $A_R^{\mathrm{shot}}$ measures correctness conditional on validity, and $A_b^{\mathrm{shot}}$ counts unresolvable shots as incorrect. The metric $T$ reports accuracy under continuous pre-measurement decoding. Values are means with 95\% bootstrap confidence intervals $[CI^{\downarrow},CI^{\uparrow}]$ over matched experimental configurations. Larger values are better for all metrics.}
    \vskip .15in
    \centering
    \setlength\tabcolsep{3mm}
    \begin{tabular}{ccccccc}
        \toprule
        K & Method & Dataset &  $R$ \;\;[$CI^{\downarrow}$, $CI^{\uparrow}$] & 
                     $A_R^{\mathrm{shot}}$ [$CI^{\downarrow}$, $CI^{\uparrow}$]\% & 
                     $A_b^{\mathrm{shot}}$ [$CI^{\downarrow}$, $CI^{\uparrow}$]\% & 
                     $T$\;\;[$CI^{\downarrow}$, $CI^{\uparrow}$]\% \\
        \midrule
    3      & Tournament &   Digits   &  $93.23$\;\;$[91.52, 94.76]$   &  $58.27$\;\;$[56.18, 60.39]$   &  $54.62$\;\;$[52.53, 56.69]$   &  $74.13$\;\;$[71.92, 76.35]$ \\
    3      &  One-hot   &   Digits   &  $57.73$\;\;$[55.84, 59.64]$   &  $60.64$\;\;$[58.37, 62.99]$   &  $38.82$\;\;$[36.33, 41.32]$   &  $74.04$\;\;$[71.84, 76.19]$ \\
    3      &   Binary   &   Digits   &  $92.14$\;\;$[90.80, 93.32]$   &  $54.50$\;\;$[52.56, 56.42]$   &  $50.41$\;\;$[48.65, 52.15]$   &  $72.32$\;\;$[69.84, 74.85]$ \\
    3      &    Gray    &   Digits   &  $90.75$\;\;$[89.49, 91.96]$   &  $54.40$\;\;$[52.39, 56.45]$   &  $49.93$\;\;$[47.95, 52.01]$   &  $73.99$\;\;$[71.80, 76.18]$ \\
    \midrule
    4      & Tournament &   Digits   &  $71.20$\;\;$[67.98, 74.30]$   &  $57.55$\;\;$[54.80, 60.30]$   &  $42.89$\;\;$[39.41, 46.37]$   &  $84.47$\;\;$[81.71, 86.99]$ \\
    4      &  One-hot   &   Digits   &  $39.28$\;\;$[35.51, 42.86]$   &  $60.25$\;\;$[56.92, 63.56]$   &  $27.41$\;\;$[24.04, 30.77]$   &  $83.76$\;\;$[80.51, 86.70]$ \\
    4      &   Binary   &   Digits   & $100.0$\;\;$[100.0, 100.0]$ &  $49.04$\;\;$[46.20, 51.78]$   &  $49.04$\;\;$[46.20, 51.78]$   &  $83.85$\;\;$[80.42, 86.94]$ \\
    4      &    Gray    &   Digits   & $100.0$\;\;$[100.0, 100.0]$ &  $25.05$\;\;$[24.45, 25.65]$   &  $25.05$\;\;$[24.45, 25.65]$   &  $25.14$\;\;$[24.32, 25.88]$ \\
    \midrule
    5      & Tournament &   Digits   &  $66.05$\;\;$[64.25, 67.83]$   &  $47.36$\;\;$[43.68, 50.81]$   &  $32.60$\;\;$[29.36, 35.77]$   &  $74.73$\;\;$[71.01, 78.19]$ \\
    5      &  One-hot   &   Digits   &  $23.83$\;\;$[21.35, 26.37]$   &  $49.65$\;\;$[45.58, 53.59]$   &  $14.86$\;\;$[12.33, 17.40]$   &  $74.32$\;\;$[70.32, 78.00]$ \\
    5      &   Binary   &   Digits   &  $78.46$\;\;$[76.71, 80.08]$   &  $39.44$\;\;$[36.39, 42.39]$   &  $31.72$\;\;$[28.73, 34.61]$   &  $72.03$\;\;$[67.37, 76.42]$ \\
    5      &    Gray    &   Digits   &  $78.41$\;\;$[76.57, 80.17]$   &  $39.76$\;\;$[36.76, 42.74]$   &  $32.01$\;\;$[28.98, 34.99]$   &  $72.38$\;\;$[67.78, 76.74]$ \\
    \midrule
    6      & Tournament &   Digits   &  $67.16$\;\;$[66.06, 68.22]$   &  $42.63$\;\;$[38.16, 46.83]$   &  $29.63$\;\;$[26.17, 32.97]$   &  $68.53$\;\;$[63.44, 73.20]$ \\
    6      &  One-hot   &   Digits   &  $15.64$\;\;$[14.32, 16.96]$   &  $41.44$\;\;$[37.01, 45.74]$   &    $8.21$\;\;$[6.53, 9.92]$    &  $66.60$\;\;$[60.82, 72.04]$ \\
    6      &   Binary   &   Digits   &  $83.19$\;\;$[82.11, 84.21]$   &  $31.02$\;\;$[28.32, 33.70]$   &  $26.23$\;\;$[23.71, 28.76]$   &  $61.72$\;\;$[56.04, 67.18]$ \\
    6      &    Gray    &   Digits   &  $83.71$\;\;$[82.50, 84.84]$   &  $31.39$\;\;$[28.69, 34.05]$   &  $26.75$\;\;$[24.17, 29.28]$   &  $62.00$\;\;$[55.88, 67.79]$ \\
    \midrule
    3      & Tournament &  Fashion   &  $97.41$\;\;$[96.16, 98.50]$   &  $60.46$\;\;$[57.96, 63.05]$   &  $59.03$\;\;$[56.51, 61.58]$   &  $75.42$\;\;$[71.67, 78.99]$ \\
    3      &  One-hot   &  Fashion   &  $56.24$\;\;$[48.95, 63.10]$   &  $60.51$\;\;$[57.18, 63.83]$   &  $41.59$\;\;$[36.79, 46.30]$   &  $74.23$\;\;$[70.30, 77.95]$ \\
    3      &   Binary   &  Fashion   &  $90.71$\;\;$[88.71, 92.67]$   &  $56.58$\;\;$[53.24, 59.94]$   &  $52.27$\;\;$[49.01, 55.48]$   &  $75.26$\;\;$[70.54, 79.47]$ \\
    3      &    Gray    &  Fashion   &  $95.11$\;\;$[93.49, 96.54]$   &  $56.90$\;\;$[53.98, 59.85]$   &  $54.49$\;\;$[51.24, 57.73]$   &  $76.16$\;\;$[72.20, 79.97]$ \\
    \midrule
    4      & Tournament &  Fashion   &  $76.36$\;\;$[72.73, 79.81]$   &  $53.28$\;\;$[51.10, 55.49]$   &  $41.95$\;\;$[38.78, 45.05]$   &  $79.57$\;\;$[77.22, 81.91]$ \\
    4      &  One-hot   &  Fashion   &  $38.02$\;\;$[34.20, 41.95]$   &  $55.16$\;\;$[52.54, 57.78]$   &  $25.02$\;\;$[21.88, 28.16]$   &  $79.15$\;\;$[76.24, 81.84]$ \\
    4      &   Binary   &  Fashion   & $100.0$\;\;$[100.0, 100.0]$ &  $45.68$\;\;$[43.53, 47.72]$   &  $45.68$\;\;$[43.53, 47.72]$   &  $79.44$\;\;$[76.15, 82.51]$ \\
    4      &    Gray    &  Fashion   & $100.0$\;\;$[100.0, 100.0]$ &  $25.22$\;\;$[24.69, 25.74]$   &  $25.22$\;\;$[24.69, 25.74]$   &  $26.74$\;\;$[25.47, 27.95]$ \\
    \midrule
    5      & Tournament &  Fashion   &  $71.43$\;\;$[69.01, 73.63]$   &  $48.26$\;\;$[44.98, 51.46]$   &  $35.73$\;\;$[32.37, 38.92]$   &  $74.88$\;\;$[71.72, 78.07]$ \\
    5      &  One-hot   &  Fashion   &  $25.07$\;\;$[22.84, 27.40]$   &  $50.78$\;\;$[47.46, 53.98]$   &  $16.59$\;\;$[14.18, 18.99]$   &  $74.79$\;\;$[71.82, 77.73]$ \\
    5      &   Binary   &  Fashion   &  $81.73$\;\;$[80.01, 83.46]$   &  $41.63$\;\;$[38.89, 44.36]$   &  $34.68$\;\;$[31.87, 37.45]$   &  $72.53$\;\;$[69.06, 75.94]$ \\
    5      &    Gray    &  Fashion   &  $81.75$\;\;$[79.99, 83.47]$   &  $42.63$\;\;$[40.05, 45.18]$   &  $35.40$\;\;$[32.72, 38.06]$   &  $74.03$\;\;$[70.91, 77.19]$ \\
    \midrule
    6      & Tournament &  Fashion   &  $68.15$\;\;$[66.76, 69.53]$   &  $41.11$\;\;$[36.84, 45.29]$   &  $29.05$\;\;$[25.50, 32.55]$   &  $67.59$\;\;$[63.34, 71.64]$ \\
    6      &  One-hot   &  Fashion   &  $15.50$\;\;$[14.20, 16.85]$   &  $41.06$\;\;$[37.17, 44.97]$   &   $8.28$\;\;$[6.66, 10.01]$    &  $66.38$\;\;$[62.09, 70.52]$ \\
    6      &   Binary   &  Fashion   &  $84.06$\;\;$[82.99, 85.12]$   &  $31.01$\;\;$[28.64, 33.38]$   &  $26.49$\;\;$[24.21, 28.77]$   &  $61.17$\;\;$[56.66, 65.66]$ \\
    6      &    Gray    &  Fashion   &  $84.25$\;\;$[83.05, 85.32]$   &  $31.62$\;\;$[29.28, 33.99]$   &  $27.13$\;\;$[24.85, 29.40]$   &  $62.91$\;\;$[58.14, 67.46]$ \\    
            \bottomrule
    \end{tabular}\medskip
    \label{tab:full_dataset}
\end{table*}

\begin{table*}[t]
\scriptsize
    \centering
    \caption{Expanded metrics aggregated by class count, encoding, and circuit. The table reports resolvability ratio $R$, shot-level resolvable accuracy $A_R^{\mathrm{shot}}$, raw bitstring shot accuracy $A_b^{\mathrm{shot}}$, and continuous top-1 simulation accuracy $T$. The first three metrics evaluate discrete measurement outcomes: $R$ measures how often a shot yields a valid prediction, $A_R^{\mathrm{shot}}$ measures correctness conditional on validity, and $A_b^{\mathrm{shot}}$ counts unresolvable shots as incorrect. The metric $T$ reports accuracy under continuous pre-measurement decoding. Values are means with 95\% bootstrap confidence intervals $[CI^{\downarrow},CI^{\uparrow}]$ over matched experimental configurations. Larger values are better for all metrics.}
    \vskip .15in
    \centering
    \setlength\tabcolsep{3mm}
    \begin{tabular}{ccccccc}
        \toprule
        K & Method & Circuit &  $R$ \;\;[$CI^{\downarrow}$, $CI^{\uparrow}$] & 
                     $A_R^{\mathrm{shot}}$ [$CI^{\downarrow}$, $CI^{\uparrow}$]\% & 
                     $A_b^{\mathrm{shot}}$ [$CI^{\downarrow}$, $CI^{\uparrow}$]\% & 
                     $T$\;\;[$CI^{\downarrow}$, $CI^{\uparrow}$]\% \\
        \midrule
    3      & Tournament &    CNN7    &  $97.89$\;\;$[96.28, 99.25]$   &  $60.55$\;\;$[56.70, 64.67]$   &  $59.28$\;\;$[55.59, 63.31]$   &  $75.63$\;\;$[70.13, 80.89]$ \\
    3      & Tournament &    CNN8    &  $97.97$\;\;$[96.35, 99.30]$   &  $60.63$\;\;$[56.68, 64.81]$   &  $59.40$\;\;$[55.63, 63.53]$   &  $75.95$\;\;$[70.48, 81.17]$ \\
    3      & Tournament &   SEL-Z    &  $89.14$\;\;$[86.55, 92.06]$   &  $58.90$\;\;$[54.94, 62.97]$   &  $53.18$\;\;$[49.41, 57.05]$   &  $74.57$\;\;$[69.30, 79.56]$ \\
    3      & Tournament &   SEL-X    &  $95.78$\;\;$[93.53, 97.70]$   &  $60.46$\;\;$[56.42, 64.86]$   &  $58.04$\;\;$[54.38, 62.08]$   &  $75.36$\;\;$[70.31, 80.03]$ \\
    3      & Tournament &   SO(4)    &  $93.28$\;\;$[91.29, 95.21]$   &  $55.27$\;\;$[51.49, 59.24]$   &  $51.93$\;\;$[48.25, 55.73]$   &  $71.39$\;\;$[66.18, 76.73]$ \\
    3      & Tournament &   SU(4)    &  $97.85$\;\;$[96.37, 99.10]$   &  $60.39$\;\;$[56.49, 64.56]$   &  $59.12$\;\;$[55.39, 63.24]$   &  $75.76$\;\;$[70.28, 81.04]$ \\
    \midrule
    4      & Tournament &    CNN7    &  $79.47$\;\;$[77.14, 81.84]$   &  $59.83$\;\;$[56.57, 63.00]$   &  $48.35$\;\;$[45.73, 50.99]$   &  $85.32$\;\;$[81.39, 88.60]$ \\
    4      & Tournament &    CNN8    &  $80.98$\;\;$[78.93, 82.94]$   &  $60.93$\;\;$[57.81, 63.91]$   &  $50.10$\;\;$[47.75, 52.55]$   &  $86.10$\;\;$[82.30, 89.34]$ \\
    4      & Tournament &   SEL-Z    &  $60.91$\;\;$[58.91, 62.94]$   &  $48.86$\;\;$[46.91, 50.90]$   &  $32.00$\;\;$[30.02, 34.14]$   &  $73.04$\;\;$[70.35, 75.63]$ \\
    4      & Tournament &   SEL-X    &  $62.26$\;\;$[59.64, 65.38]$   &  $47.82$\;\;$[46.45, 49.15]$   &  $30.89$\;\;$[29.60, 32.23]$   &  $79.23$\;\;$[75.21, 83.01]$ \\
    4      & Tournament &   SO(4)    &  $76.27$\;\;$[73.61, 78.81]$   &  $52.91$\;\;$[50.17, 55.56]$   &  $41.08$\;\;$[38.79, 43.47]$   &  $81.97$\;\;$[78.08, 85.53]$ \\
    4      & Tournament &   SU(4)    &  $82.77$\;\;$[80.56, 84.91]$   &  $62.14$\;\;$[58.87, 65.37]$   &  $52.12$\;\;$[49.73, 54.56]$   &  $86.46$\;\;$[82.40, 90.01]$ \\
    \midrule
    5      & Tournament &    CNN7    &  $71.86$\;\;$[69.82, 73.82]$   &  $54.09$\;\;$[52.11, 56.22]$   &  $39.85$\;\;$[37.66, 42.17]$   &  $80.03$\;\;$[77.00, 83.28]$ \\
    5      & Tournament &    CNN8    &  $72.54$\;\;$[70.17, 74.79]$   &  $55.07$\;\;$[52.98, 57.42]$   &  $40.96$\;\;$[38.63, 43.62]$   &  $81.03$\;\;$[78.10, 84.18]$ \\
    5      & Tournament &   SEL-Z    &  $66.26$\;\;$[63.05, 69.77]$   &  $42.22$\;\;$[39.73, 44.59]$   &  $29.02$\;\;$[26.87, 30.84]$   &  $65.19$\;\;$[62.52, 68.41]$ \\
    5      & Tournament &   SEL-X    &  $59.12$\;\;$[58.57, 59.76]$   &  $31.66$\;\;$[30.24, 33.03]$   &  $19.02$\;\;$[18.06, 19.92]$   &  $61.84$\;\;$[59.27, 64.55]$ \\
    5      & Tournament &   SO(4)    &  $67.44$\;\;$[65.72, 69.21]$   &  $46.89$\;\;$[44.37, 49.15]$   &  $32.36$\;\;$[30.08, 34.61]$   &  $78.34$\;\;$[75.14, 81.75]$ \\
    5      & Tournament &   SU(4)    &  $75.22$\;\;$[73.10, 77.17]$   &  $56.93$\;\;$[54.69, 59.08]$   &  $43.80$\;\;$[41.62, 45.95]$   &  $82.41$\;\;$[79.83, 85.08]$ \\
    \midrule
    6      & Tournament &    CNN7    &  $69.73$\;\;$[68.96, 70.69]$   &  $48.95$\;\;$[45.43, 53.18]$   &  $35.00$\;\;$[32.21, 38.47]$   &  $75.29$\;\;$[72.28, 78.62]$ \\
    6      & Tournament &    CNN8    &  $70.15$\;\;$[68.88, 71.56]$   &  $50.18$\;\;$[46.41, 54.59]$   &  $36.23$\;\;$[32.99, 40.08]$   &  $76.32$\;\;$[73.23, 79.67]$ \\
    6      & Tournament &   SEL-Z    &  $65.23$\;\;$[64.15, 66.49]$   &  $36.93$\;\;$[34.21, 40.11]$   &  $24.65$\;\;$[22.56, 27.21]$   &  $59.99$\;\;$[56.22, 63.94]$ \\
    6      & Tournament &   SEL-X    &  $62.56$\;\;$[62.49, 62.62]$   &  $21.15$\;\;$[20.80, 21.54]$   &  $13.29$\;\;$[13.06, 13.55]$   &  $46.26$\;\;$[42.84, 50.11]$ \\
    6      & Tournament &   SO(4)    &  $66.79$\;\;$[66.22, 67.40]$   &  $41.40$\;\;$[38.67, 44.58]$   &  $28.13$\;\;$[26.07, 30.50]$   &  $72.92$\;\;$[69.58, 76.49]$ \\
    6      & Tournament &   SU(4)    &  $71.49$\;\;$[70.20, 72.80]$   &  $52.62$\;\;$[48.91, 56.75]$   &  $38.75$\;\;$[35.57, 42.37]$   &  $77.57$\;\;$[74.72, 80.61]$ \\
    \midrule
    3      &  One-hot   &    CNN7    &  $56.24$\;\;$[45.75, 64.83]$   &  $61.35$\;\;$[55.81, 66.83]$   &  $42.34$\;\;$[36.17, 48.56]$   &  $74.70$\;\;$[68.73, 79.92]$ \\
    3      &  One-hot   &    CNN8    &  $56.62$\;\;$[45.62, 65.60]$   &  $60.81$\;\;$[55.40, 66.18]$   &  $42.31$\;\;$[35.87, 48.62]$   &  $74.45$\;\;$[68.25, 79.97]$ \\
    3      &  One-hot   &   SEL-Z    &  $65.65$\;\;$[58.98, 72.41]$   &  $62.10$\;\;$[57.65, 66.61]$   &  $42.99$\;\;$[36.07, 49.99]$   &  $74.73$\;\;$[69.50, 79.61]$ \\
    3      &  One-hot   &   SEL-X    &  $56.55$\;\;$[45.83, 65.43]$   &  $61.57$\;\;$[56.75, 66.41]$   &  $41.55$\;\;$[34.85, 48.18]$   &  $74.15$\;\;$[68.02, 79.54]$ \\
    3      &  One-hot   &   SO(4)    &  $50.65$\;\;$[47.07, 53.95]$   &  $56.75$\;\;$[53.02, 60.44]$   &  $29.97$\;\;$[27.41, 32.66]$   &  $71.41$\;\;$[66.19, 76.71]$ \\
    3      &  One-hot   &   SU(4)    &  $56.19$\;\;$[45.42, 65.26]$   &  $60.87$\;\;$[55.37, 66.33]$   &  $42.06$\;\;$[35.64, 48.52]$   &  $75.37$\;\;$[70.31, 80.12]$ \\
    \midrule
    4      &  One-hot   &    CNN7    &  $44.05$\;\;$[40.61, 47.14]$   &  $63.17$\;\;$[59.37, 66.83]$   &  $31.45$\;\;$[28.85, 33.98]$   &  $85.48$\;\;$[81.76, 88.75]$ \\
    4      &  One-hot   &    CNN8    &  $45.76$\;\;$[40.50, 50.24]$   &  $64.26$\;\;$[60.05, 68.21]$   &  $33.56$\;\;$[29.84, 37.06]$   &  $86.07$\;\;$[82.58, 89.13]$ \\
    4      &  One-hot   &   SEL-Z    &  $27.15$\;\;$[23.07, 31.49]$   &  $48.46$\;\;$[46.58, 50.45]$   &  $16.34$\;\;$[14.18, 18.77]$   &  $68.80$\;\;$[64.82, 72.50]$ \\
    4      &  One-hot   &   SEL-X    &  $27.82$\;\;$[25.94, 29.75]$   &  $49.83$\;\;$[46.66, 53.12]$   &  $16.32$\;\;$[14.28, 18.45]$   &  $79.29$\;\;$[74.32, 83.73]$ \\
    4      &  One-hot   &   SO(4)    &  $38.19$\;\;$[34.13, 42.29]$   &  $55.69$\;\;$[52.27, 59.01]$   &  $24.18$\;\;$[21.28, 27.29]$   &  $82.94$\;\;$[79.06, 86.47]$ \\
    4      &  One-hot   &   SU(4)    &  $48.95$\;\;$[44.36, 53.07]$   &  $64.82$\;\;$[60.87, 68.55]$   &  $35.43$\;\;$[31.66, 39.02]$   &  $86.15$\;\;$[82.55, 89.32]$ \\
    \midrule
    5      &  One-hot   &    CNN7    &  $27.79$\;\;$[25.26, 30.08]$   &  $56.24$\;\;$[53.58, 58.74]$   &  $19.33$\;\;$[17.18, 21.13]$   &  $79.59$\;\;$[77.11, 82.31]$ \\
    5      &  One-hot   &    CNN8    &  $29.44$\;\;$[27.29, 31.70]$   &  $58.03$\;\;$[55.46, 60.75]$   &  $21.04$\;\;$[19.39, 23.22]$   &  $80.86$\;\;$[78.07, 83.80]$ \\
    5      &  One-hot   &   SEL-Z    &  $17.16$\;\;$[15.66, 19.10]$   &  $38.10$\;\;$[34.12, 42.20]$   &   $8.48$\;\;$[6.61, 10.63]$    &  $60.82$\;\;$[56.61, 65.30]$ \\
    5      &  One-hot   &   SEL-X    &  $16.60$\;\;$[15.58, 17.61]$   &  $38.67$\;\;$[35.17, 42.02]$   &    $7.33$\;\;$[6.09, 8.63]$    &  $66.23$\;\;$[62.93, 69.63]$ \\
    5      &  One-hot   &   SO(4)    &  $24.37$\;\;$[22.82, 25.85]$   &  $50.21$\;\;$[47.95, 52.48]$   &  $14.82$\;\;$[13.18, 16.36]$   &  $77.32$\;\;$[75.11, 79.78]$ \\
    5      &  One-hot   &   SU(4)    &  $31.37$\;\;$[28.23, 34.82]$   &  $60.04$\;\;$[56.96, 63.10]$   &  $23.36$\;\;$[20.45, 26.36]$   &  $82.52$\;\;$[79.95, 85.19]$ \\
    \midrule
    6      &  One-hot   &    CNN7    &  $16.46$\;\;$[15.56, 17.29]$   &  $46.76$\;\;$[43.45, 49.86]$   &   $9.59$\;\;$[8.32, 10.80]$    &  $74.24$\;\;$[70.44, 78.24]$ \\
    6      &  One-hot   &    CNN8    &  $18.10$\;\;$[16.90, 19.70]$   &  $49.58$\;\;$[45.95, 53.47]$   &  $11.88$\;\;$[10.19, 13.82]$   &  $75.78$\;\;$[72.65, 79.42]$ \\
    6      &  One-hot   &   SEL-Z    &  $12.28$\;\;$[11.73, 12.91]$   &  $31.13$\;\;$[28.97, 33.24]$   &    $3.91$\;\;$[3.39, 4.48]$    &  $49.35$\;\;$[44.30, 54.90]$ \\
    6      &  One-hot   &   SEL-X    &  $10.73$\;\;$[10.60, 10.87]$   &  $24.04$\;\;$[22.96, 25.11]$   &    $2.27$\;\;$[2.12, 2.44]$    &  $49.02$\;\;$[46.80, 52.14]$ \\
    6      &  One-hot   &   SO(4)    &  $15.36$\;\;$[14.56, 16.15]$   &  $41.68$\;\;$[39.39, 43.83]$   &    $7.48$\;\;$[6.70, 8.28]$    &  $72.20$\;\;$[68.72, 75.76]$ \\
    6      &  One-hot   &   SU(4)    &  $20.52$\;\;$[19.05, 22.03]$   &  $54.31$\;\;$[50.96, 57.97]$   &  $14.34$\;\;$[12.55, 16.23]$   &  $78.33$\;\;$[75.15, 81.60]$ \\
        \bottomrule
    \end{tabular}\medskip
    
    \label{tab:full_circuit_1}
\end{table*}

\begin{table*}[t]
\scriptsize
    \centering
    \caption{(Continued) Expanded metrics aggregated by class count, encoding, and circuit. The table reports resolvability ratio $R$, shot-level resolvable accuracy $A_R^{\mathrm{shot}}$, raw bitstring shot accuracy $A_b^{\mathrm{shot}}$, and continuous top-1 simulation accuracy $T$. The first three metrics evaluate discrete measurement outcomes: $R$ measures how often a shot yields a valid prediction, $A_R^{\mathrm{shot}}$ measures correctness conditional on validity, and $A_b^{\mathrm{shot}}$ counts unresolvable shots as incorrect. The metric $T$ reports accuracy under continuous pre-measurement decoding. Values are means with 95\% bootstrap confidence intervals $[CI^{\downarrow},CI^{\uparrow}]$ over matched experimental configurations. Larger values are better for all metrics.}
    \vskip .15in
    \centering
    \setlength\tabcolsep{3mm}
    \begin{tabular}{ccccccc}
        \toprule
        K & Method & Circuit &  $R$ \;\;[$CI^{\downarrow}$, $CI^{\uparrow}$] & 
                     $A_R^{\mathrm{shot}}$ [$CI^{\downarrow}$, $CI^{\uparrow}$]\% & 
                     $A_b^{\mathrm{shot}}$ [$CI^{\downarrow}$, $CI^{\uparrow}$]\% & 
                     $T$\;\;[$CI^{\downarrow}$, $CI^{\uparrow}$]\% \\
        \midrule
    3      &   Binary   &    CNN7    &  $92.33$\;\;$[90.31, 94.44]$   &  $56.83$\;\;$[52.21, 61.66]$   &  $52.89$\;\;$[48.81, 57.06]$   &  $76.10$\;\;$[70.81, 81.44]$ \\
    3      &   Binary   &    CNN8    &  $92.32$\;\;$[89.71, 94.92]$   &  $56.90$\;\;$[52.46, 61.70]$   &  $53.01$\;\;$[48.91, 57.35]$   &  $75.11$\;\;$[69.67, 80.71]$ \\
    3      &   Binary   &   SEL-Z    &  $88.38$\;\;$[84.42, 91.83]$   &  $54.44$\;\;$[49.26, 59.46]$   &  $49.10$\;\;$[43.94, 54.10]$   &  $71.31$\;\;$[61.22, 78.92]$ \\
    3      &   Binary   &   SEL-X    &  $92.20$\;\;$[89.87, 94.38]$   &  $56.35$\;\;$[51.82, 61.00]$   &  $52.40$\;\;$[48.17, 56.81]$   &  $75.61$\;\;$[70.17, 81.24]$ \\
    3      &   Binary   &   SO(4)    &  $91.29$\;\;$[89.09, 93.54]$   &  $52.10$\;\;$[48.00, 56.66]$   &  $47.97$\;\;$[44.02, 52.22]$   &  $70.56$\;\;$[64.70, 76.62]$ \\
    3      &   Binary   &   SU(4)    &  $92.04$\;\;$[88.42, 95.16]$   &  $56.61$\;\;$[51.86, 61.39]$   &  $52.67$\;\;$[47.91, 57.02]$   &  $74.05$\;\;$[68.49, 79.67]$ \\
    \midrule
    4      &   Binary   &    CNN7    & $100.0$\;\;$[100.0, 100.0]$ &  $52.01$\;\;$[49.40, 54.60]$   &  $52.01$\;\;$[49.40, 54.60]$   &  $85.75$\;\;$[81.82, 89.20]$ \\
    4      &   Binary   &    CNN8    & $100.0$\;\;$[100.0, 100.0]$ &  $52.28$\;\;$[49.86, 54.47]$   &  $52.28$\;\;$[49.86, 54.47]$   &  $86.64$\;\;$[82.69, 90.05]$ \\
    4      &   Binary   &   SEL-Z    & $100.0$\;\;$[100.0, 100.0]$ &  $35.70$\;\;$[34.90, 36.57]$   &  $35.70$\;\;$[34.90, 36.57]$   &  $66.77$\;\;$[62.74, 71.11]$ \\
    4      &   Binary   &   SEL-X    & $100.0$\;\;$[100.0, 100.0]$ &  $44.61$\;\;$[43.23, 46.15]$   &  $44.61$\;\;$[43.23, 46.15]$   &  $80.63$\;\;$[77.16, 83.93]$ \\
    4      &   Binary   &   SO(4)    & $100.0$\;\;$[100.0, 100.0]$ &  $46.43$\;\;$[44.38, 48.33]$   &  $46.43$\;\;$[44.38, 48.33]$   &  $83.28$\;\;$[79.50, 86.58]$ \\
    4      &   Binary   &   SU(4)    & $100.0$\;\;$[100.0, 100.0]$ &  $53.12$\;\;$[49.92, 56.51]$   &  $53.12$\;\;$[49.92, 56.51]$   &  $86.81$\;\;$[82.08, 90.67]$ \\
    \midrule
    5      &   Binary   &    CNN7    &  $82.54$\;\;$[80.72, 84.62]$   &  $45.09$\;\;$[42.72, 47.39]$   &  $37.64$\;\;$[35.33, 39.82]$   &  $77.61$\;\;$[74.42, 80.90]$ \\
    5      &   Binary   &    CNN8    &  $83.31$\;\;$[81.32, 85.60]$   &  $47.74$\;\;$[45.47, 50.24]$   &  $40.34$\;\;$[37.97, 42.76]$   &  $80.60$\;\;$[77.82, 83.49]$ \\
    5      &   Binary   &   SEL-Z    &  $75.49$\;\;$[74.24, 76.75]$   &  $31.02$\;\;$[29.73, 32.37]$   &  $23.76$\;\;$[22.49, 25.09]$   &  $58.46$\;\;$[55.62, 60.87]$ \\
    5      &   Binary   &   SEL-X    &  $73.97$\;\;$[71.36, 76.33]$   &  $30.86$\;\;$[28.79, 33.06]$   &  $23.24$\;\;$[21.10, 25.48]$   &  $58.44$\;\;$[54.37, 63.50]$ \\
    5      &   Binary   &   SO(4)    &  $81.19$\;\;$[79.68, 83.14]$   &  $40.22$\;\;$[38.41, 41.93]$   &  $33.07$\;\;$[31.27, 34.89]$   &  $76.26$\;\;$[73.14, 79.50]$ \\
    5      &   Binary   &   SU(4)    &  $84.07$\;\;$[82.28, 86.19]$   &  $48.26$\;\;$[45.51, 50.84]$   &  $41.14$\;\;$[38.39, 43.67]$   &  $82.31$\;\;$[79.71, 84.77]$ \\
    
    \midrule
    6      &   Binary   &    CNN7    &  $85.64$\;\;$[84.69, 86.66]$   &  $34.51$\;\;$[32.64, 36.64]$   &  $29.85$\;\;$[28.08, 31.89]$   &  $67.95$\;\;$[64.20, 72.11]$ \\
    6      &   Binary   &    CNN8    &  $85.42$\;\;$[84.11, 86.85]$   &  $37.23$\;\;$[34.88, 39.91]$   &  $32.15$\;\;$[29.86, 34.69]$   &  $72.47$\;\;$[68.88, 76.38]$ \\
    6      &   Binary   &   SEL-Z    &  $82.45$\;\;$[81.42, 83.58]$   &  $23.60$\;\;$[22.52, 24.91]$   &  $19.65$\;\;$[18.60, 20.91]$   &  $46.56$\;\;$[42.70, 50.75]$ \\
    6      &   Binary   &   SEL-X    &  $78.80$\;\;$[77.87, 79.77]$   &  $21.58$\;\;$[20.89, 22.23]$   &  $17.14$\;\;$[16.70, 17.62]$   &  $41.65$\;\;$[39.84, 43.40]$ \\
    6      &   Binary   &   SO(4)    &  $83.46$\;\;$[83.10, 83.85]$   &  $30.80$\;\;$[29.30, 32.68]$   &  $25.97$\;\;$[24.69, 27.57]$   &  $65.15$\;\;$[60.85, 70.06]$ \\
    6      &   Binary   &   SU(4)    &  $85.99$\;\;$[85.06, 86.98]$   &  $38.36$\;\;$[36.43, 40.63]$   &  $33.38$\;\;$[31.57, 35.59]$   &  $74.87$\;\;$[71.04, 78.96]$ \\
    \midrule
    3      &    Gray    &    CNN7    &  $93.99$\;\;$[90.98, 96.75]$   &  $56.32$\;\;$[51.90, 61.05]$   &  $53.51$\;\;$[48.39, 59.00]$   &  $75.81$\;\;$[70.42, 81.07]$ \\
    3      &    Gray    &    CNN8    &  $94.33$\;\;$[91.88, 96.66]$   &  $56.99$\;\;$[52.65, 61.75]$   &  $54.25$\;\;$[49.47, 59.56]$   &  $75.74$\;\;$[70.15, 81.11]$ \\
    3      &    Gray    &   SEL-Z    &  $90.78$\;\;$[88.88, 92.80]$   &  $55.83$\;\;$[51.71, 59.89]$   &  $51.00$\;\;$[47.23, 54.66]$   &  $75.28$\;\;$[69.36, 81.00]$ \\
    3      &    Gray    &   SEL-X    &  $92.75$\;\;$[90.01, 95.33]$   &  $56.54$\;\;$[52.48, 60.84]$   &  $52.92$\;\;$[48.55, 57.59]$   &  $76.18$\;\;$[70.80, 81.40]$ \\
    3      &    Gray    &   SO(4)    &  $91.58$\;\;$[88.95, 94.33]$   &  $51.66$\;\;$[47.86, 55.82]$   &  $47.78$\;\;$[43.62, 52.48]$   &  $71.82$\;\;$[66.27, 77.62]$ \\
    3      &    Gray    &   SU(4)    &  $94.13$\;\;$[90.89, 97.01]$   &  $56.55$\;\;$[52.00, 61.49]$   &  $53.84$\;\;$[48.70, 59.34]$   &  $75.61$\;\;$[70.18, 81.01]$ \\
    \midrule
    4      &    Gray    &    CNN7    & $100.0$\;\;$[100.0, 100.0]$ &  $24.94$\;\;$[23.92, 25.90]$   &  $24.94$\;\;$[23.92, 25.90]$   &  $25.72$\;\;$[24.34, 27.19]$ \\
    4      &    Gray    &    CNN8    & $100.0$\;\;$[100.0, 100.0]$ &  $25.18$\;\;$[24.00, 26.26]$   &  $25.18$\;\;$[24.00, 26.26]$   &  $25.83$\;\;$[24.18, 27.47]$ \\
    4      &    Gray    &   SEL-Z    & $100.0$\;\;$[100.0, 100.0]$ &  $25.31$\;\;$[24.31, 26.39]$   &  $25.31$\;\;$[24.31, 26.39]$   &  $26.06$\;\;$[23.40, 28.54]$ \\
    4      &    Gray    &   SEL-X    & $100.0$\;\;$[100.0, 100.0]$ &  $25.15$\;\;$[24.35, 25.92]$   &  $25.15$\;\;$[24.35, 25.92]$   &  $26.02$\;\;$[24.43, 27.69]$ \\
    4      &    Gray    &   SO(4)    & $100.0$\;\;$[100.0, 100.0]$ &  $25.11$\;\;$[24.15, 26.03]$   &  $25.11$\;\;$[24.15, 26.03]$   &  $25.90$\;\;$[24.06, 27.65]$ \\
    4      &    Gray    &   SU(4)    & $100.0$\;\;$[100.0, 100.0]$ &  $25.14$\;\;$[24.20, 26.03]$   &  $25.14$\;\;$[24.20, 26.03]$   &  $26.10$\;\;$[24.27, 27.86]$ \\
    \midrule
    5      &    Gray    &    CNN7    &  $82.19$\;\;$[80.03, 84.79]$   &  $45.61$\;\;$[43.61, 47.72]$   &  $37.96$\;\;$[35.68, 40.24]$   &  $78.74$\;\;$[75.75, 81.90]$ \\
    5      &    Gray    &    CNN8    &  $83.47$\;\;$[81.95, 85.36]$   &  $46.90$\;\;$[44.37, 49.67]$   &  $39.58$\;\;$[37.06, 42.30]$   &  $80.90$\;\;$[77.89, 84.08]$ \\
    5      &    Gray    &   SEL-Z    &  $75.27$\;\;$[73.62, 76.78]$   &  $31.78$\;\;$[29.68, 33.99]$   &  $24.26$\;\;$[22.28, 26.30]$   &  $60.06$\;\;$[54.63, 65.02]$ \\
    5      &    Gray    &   SEL-X    &  $74.18$\;\;$[71.38, 76.95]$   &  $32.84$\;\;$[30.41, 35.46]$   &  $24.85$\;\;$[22.41, 27.41]$   &  $61.78$\;\;$[57.45, 66.09]$ \\
    5      &    Gray    &   SO(4)    &  $80.70$\;\;$[79.15, 82.82]$   &  $40.62$\;\;$[38.18, 43.09]$   &  $33.23$\;\;$[30.74, 35.75]$   &  $75.17$\;\;$[72.04, 78.68]$ \\
    5      &    Gray    &   SU(4)    &  $84.68$\;\;$[83.18, 86.53]$   &  $49.42$\;\;$[46.95, 52.23]$   &  $42.36$\;\;$[39.74, 45.34]$   &  $82.59$\;\;$[79.47, 85.70]$ \\
    \midrule
    6      &    Gray    &    CNN7    &  $85.20$\;\;$[84.46, 85.97]$   &  $35.51$\;\;$[33.30, 37.81]$   &  $30.66$\;\;$[28.69, 32.66]$   &  $70.07$\;\;$[65.65, 74.43]$ \\
    6      &    Gray    &    CNN8    &  $86.99$\;\;$[86.17, 87.80]$   &  $37.41$\;\;$[35.47, 39.65]$   &  $32.89$\;\;$[31.22, 34.85]$   &  $74.31$\;\;$[71.08, 77.80]$ \\
    6      &    Gray    &   SEL-Z    &  $82.33$\;\;$[81.13, 83.43]$   &  $24.30$\;\;$[23.41, 25.15]$   &  $20.17$\;\;$[19.46, 20.86]$   &  $49.15$\;\;$[46.08, 51.68]$ \\
    6      &    Gray    &   SEL-X    &  $78.63$\;\;$[77.40, 79.75]$   &  $22.06$\;\;$[20.96, 23.28]$   &  $17.53$\;\;$[16.58, 18.62]$   &  $39.52$\;\;$[35.01, 44.46]$ \\
    6      &    Gray    &   SO(4)    &  $84.05$\;\;$[83.40, 84.68]$   &  $30.84$\;\;$[29.48, 32.36]$   &  $26.27$\;\;$[25.08, 27.62]$   &  $64.99$\;\;$[62.16, 68.49]$ \\
    6      &    Gray    &   SU(4)    &  $86.69$\;\;$[86.03, 87.29]$   &  $38.93$\;\;$[36.93, 41.17]$   &  $34.14$\;\;$[32.29, 36.17]$   &  $76.69$\;\;$[73.55, 80.12]$ \\
        \bottomrule
    \end{tabular}\medskip
    
    \label{tab:full_circuit_2}
\end{table*}
\begin{table*}[t]
\scriptsize
    \centering
    \caption{Expanded metrics aggregated by class count, encoding, dataset, and circuit. The table reports resolvability ratio $R$, shot-level resolvable accuracy $A_R^{\mathrm{shot}}$, raw bitstring shot accuracy $A_b^{\mathrm{shot}}$, and continuous top-1 simulation accuracy $T$. The first three metrics evaluate discrete measurement outcomes: $R$ measures how often a shot yields a valid prediction, $A_R^{\mathrm{shot}}$ measures correctness conditional on validity, and $A_b^{\mathrm{shot}}$ counts unresolvable shots as incorrect. The metric $T$ reports accuracy under continuous pre-measurement decoding. Values are means with 95\% bootstrap confidence intervals $[CI^{\downarrow},CI^{\uparrow}]$ over matched experimental configurations. Larger values are better for all metrics.}
    \vskip .15in
    \centering
    \setlength\tabcolsep{2mm}
    \begin{tabular}{cccccccc}
        \toprule
        K & Method & Dataset & Circuit &  $R$ \;\;[$CI^{\downarrow}$, $CI^{\uparrow}$] & 
                     $A_R^{\mathrm{shot}}$ [$CI^{\downarrow}$, $CI^{\uparrow}$]\% & 
                     $A_b^{\mathrm{shot}}$ [$CI^{\downarrow}$, $CI^{\uparrow}$]\% & 
                     $T$\;\;[$CI^{\downarrow}$, $CI^{\uparrow}$]\% \\
        \midrule    
    3      & Tournament &   Digits   &    CNN7    &  $96.13$\;\;$[94.17, 98.08]$   &  $59.93$\;\;$[55.28, 64.67]$   &  $57.61$\;\;$[53.51, 61.66]$   &  $75.57$\;\;$[69.97, 81.19]$ \\
    3      & Tournament &   Digits   &    CNN8    &  $96.33$\;\;$[94.25, 98.41]$   &  $59.88$\;\;$[55.04, 64.48]$   &  $57.66$\;\;$[53.54, 61.71]$   &  $75.97$\;\;$[71.16, 80.98]$ \\
    3      & Tournament &   Digits   &   SEL-Z    &  $85.55$\;\;$[84.42, 86.65]$   &  $58.23$\;\;$[53.56, 63.05]$   &  $50.88$\;\;$[46.44, 55.10]$   &  $74.54$\;\;$[68.92, 79.57]$ \\
    3      & Tournament &   Digits   &   SEL-X    &  $93.45$\;\;$[90.64, 96.25]$   &  $60.03$\;\;$[54.91, 65.86]$   &  $56.21$\;\;$[52.41, 60.72]$   &  $75.66$\;\;$[71.43, 80.11]$ \\
    3      & Tournament &   Digits   &   SO(4)    &  $91.77$\;\;$[89.12, 94.14]$   &  $51.94$\;\;$[48.06, 55.20]$   &  $48.03$\;\;$[44.76, 50.93]$   &  $67.71$\;\;$[63.70, 71.01]$ \\
    3      & Tournament &   Digits   &   SU(4)    &  $96.14$\;\;$[94.43, 97.85]$   &  $59.62$\;\;$[55.23, 64.31]$   &  $57.36$\;\;$[53.58, 61.64]$   &  $75.35$\;\;$[69.74, 80.88]$ \\
    \midrule
    4      & Tournament &   Digits   &    CNN7    &  $76.99$\;\;$[74.77, 79.43]$   &  $62.78$\;\;$[58.74, 66.34]$   &  $49.54$\;\;$[44.96, 53.66]$   &  $88.66$\;\;$[86.37, 90.28]$ \\
    4      & Tournament &   Digits   &    CNN8    &  $78.43$\;\;$[76.83, 81.03]$   &  $63.85$\;\;$[60.35, 67.07]$   &  $51.18$\;\;$[47.59, 54.86]$   &  $89.46$\;\;$[86.58, 91.35]$ \\
    4      & Tournament &   Digits   &   SEL-Z    &  $60.16$\;\;$[57.46, 63.11]$   &  $50.78$\;\;$[48.51, 53.05]$   &  $33.49$\;\;$[30.57, 36.41]$   &  $72.78$\;\;$[69.64, 75.92]$ \\
    4      & Tournament &   Digits   &   SEL-X    &  $59.31$\;\;$[57.79, 61.20]$   &  $48.87$\;\;$[46.72, 50.55]$   &  $30.37$\;\;$[28.42, 32.80]$   &  $81.44$\;\;$[74.58, 86.81]$ \\
    4      & Tournament &   Digits   &   SO(4)    &  $72.79$\;\;$[70.34, 75.40]$   &  $53.65$\;\;$[49.60, 56.94]$   &  $39.80$\;\;$[36.67, 42.35]$   &  $84.36$\;\;$[79.39, 88.20]$ \\
    4      & Tournament &   Digits   &   SU(4)    &  $79.50$\;\;$[78.26, 80.74]$   &  $65.36$\;\;$[61.20, 69.15]$   &  $52.97$\;\;$[49.28, 56.67]$   &  $90.12$\;\;$[87.11, 92.27]$ \\
    \midrule
    5      & Tournament &   Digits   &    CNN7    &  $69.37$\;\;$[67.54, 71.42]$   &  $53.99$\;\;$[51.58, 56.21]$   &  $38.45$\;\;$[35.83, 41.21]$   &  $80.98$\;\;$[78.84, 82.95]$ \\
    5      & Tournament &   Digits   &    CNN8    &  $69.28$\;\;$[67.52, 71.42]$   &  $55.05$\;\;$[52.71, 57.97]$   &  $39.17$\;\;$[36.48, 42.40]$   &  $82.00$\;\;$[79.63, 83.88]$ \\
    5      & Tournament &   Digits   &   SEL-Z    &  $61.79$\;\;$[60.73, 62.85]$   &  $42.52$\;\;$[38.65, 45.75]$   &  $27.56$\;\;$[24.23, 30.37]$   &  $64.02$\;\;$[62.52, 65.98]$ \\
    5      & Tournament &   Digits   &   SEL-X    &  $58.87$\;\;$[58.35, 59.42]$   &  $30.35$\;\;$[28.60, 32.23]$   &  $18.22$\;\;$[16.88, 19.59]$   &  $59.34$\;\;$[57.19, 62.40]$ \\
    5      & Tournament &   Digits   &   SO(4)    &  $64.77$\;\;$[64.10, 65.38]$   &  $44.98$\;\;$[41.27, 47.56]$   &  $29.76$\;\;$[27.14, 31.71]$   &  $78.17$\;\;$[74.08, 81.38]$ \\
    5      & Tournament &   Digits   &   SU(4)    &  $72.24$\;\;$[70.61, 73.47]$   &  $57.28$\;\;$[54.62, 59.47]$   &  $42.45$\;\;$[39.57, 44.81]$   &  $83.89$\;\;$[81.87, 85.56]$ \\
    \midrule
    6      & Tournament &   Digits   &    CNN7    &  $68.83$\;\;$[68.31, 69.40]$   &  $49.94$\;\;$[45.84, 54.04]$   &  $35.26$\;\;$[32.04, 38.47]$   &  $76.47$\;\;$[72.59, 80.50]$ \\
    6      & Tournament &   Digits   &    CNN8    &  $69.26$\;\;$[67.57, 70.79]$   &  $51.18$\;\;$[46.04, 56.31]$   &  $36.44$\;\;$[32.09, 40.79]$   &  $78.02$\;\;$[74.20, 81.83]$ \\
    6      & Tournament &   Digits   &   SEL-Z    &  $64.87$\;\;$[64.32, 65.43]$   &  $37.79$\;\;$[34.75, 41.30]$   &  $25.09$\;\;$[23.07, 27.49]$   &  $60.42$\;\;$[54.95, 65.08]$ \\
    6      & Tournament &   Digits   &   SEL-X    &  $62.52$\;\;$[62.46, 62.59]$   &  $20.81$\;\;$[20.46, 21.15]$   &  $13.07$\;\;$[12.84, 13.31]$   &  $43.43$\;\;$[40.66, 46.20]$ \\
    6      & Tournament &   Digits   &   SO(4)    &  $66.73$\;\;$[66.06, 67.52]$   &  $41.61$\;\;$[37.90, 45.96]$   &  $28.26$\;\;$[25.50, 31.68]$   &  $73.97$\;\;$[69.16, 78.79]$ \\
    6      & Tournament &   Digits   &   SU(4)    &  $70.73$\;\;$[69.30, 72.00]$   &  $54.45$\;\;$[50.94, 58.15]$   &  $39.66$\;\;$[36.45, 43.07]$   &  $78.88$\;\;$[75.51, 82.21]$ \\
    \midrule
    3      & Tournament &  Fashion   &    CNN7    &  $99.66$\;\;$[99.54, 99.78]$   &  $61.16$\;\;$[55.15, 67.20]$   &  $60.95$\;\;$[54.98, 66.93]$   &  $75.69$\;\;$[66.67, 84.72]$ \\
    3      & Tournament &  Fashion   &    CNN8    &  $99.61$\;\;$[99.41, 99.81]$   &  $61.38$\;\;$[55.22, 67.61]$   &  $61.14$\;\;$[55.10, 67.25]$   &  $75.93$\;\;$[66.70, 85.17]$ \\
    3      & Tournament &  Fashion   &   SEL-Z    &  $92.74$\;\;$[90.06, 96.05]$   &  $59.56$\;\;$[53.36, 65.75]$   &  $55.48$\;\;$[50.05, 60.91]$   &  $74.60$\;\;$[65.85, 83.04]$ \\
    3      & Tournament &  Fashion   &   SEL-X    &  $98.10$\;\;$[97.06, 98.99]$   &  $60.89$\;\;$[54.85, 67.13]$   &  $59.87$\;\;$[54.24, 65.52]$   &  $75.07$\;\;$[66.20, 83.45]$ \\
    3      & Tournament &  Fashion   &   SO(4)    &  $94.80$\;\;$[92.63, 96.97]$   &  $58.59$\;\;$[53.30, 63.89]$   &  $55.83$\;\;$[51.10, 60.56]$   &  $75.06$\;\;$[66.14, 83.58]$ \\
    3      & Tournament &  Fashion   &   SU(4)    &  $99.56$\;\;$[99.34, 99.69]$   &  $61.15$\;\;$[55.12, 67.16]$   &  $60.89$\;\;$[54.88, 66.86]$   &  $76.17$\;\;$[67.25, 85.08]$ \\
    \midrule
    4      & Tournament &  Fashion   &    CNN7    &  $81.94$\;\;$[79.29, 84.63]$   &  $56.88$\;\;$[53.23, 60.53]$   &  $47.16$\;\;$[44.72, 49.93]$   &  $81.99$\;\;$[76.69, 87.30]$ \\
    4      & Tournament &  Fashion   &    CNN8    &  $83.53$\;\;$[82.39, 84.56]$   &  $58.00$\;\;$[54.58, 61.42]$   &  $49.03$\;\;$[46.51, 51.54]$   &  $82.75$\;\;$[77.73, 87.76]$ \\
    4      & Tournament &  Fashion   &   SEL-Z    &  $61.66$\;\;$[58.87, 64.35]$   &  $46.94$\;\;$[45.03, 49.61]$   &  $30.50$\;\;$[28.67, 33.24]$   &  $73.29$\;\;$[69.12, 77.45]$ \\
    4      & Tournament &  Fashion   &   SEL-X    &  $65.21$\;\;$[61.59, 69.41]$   &  $46.76$\;\;$[45.49, 47.86]$   &  $31.40$\;\;$[30.09, 32.71]$   &  $77.03$\;\;$[73.05, 80.91]$ \\
    4      & Tournament &  Fashion   &   SO(4)    &  $79.76$\;\;$[77.82, 81.14]$   &  $52.17$\;\;$[48.87, 56.12]$   &  $42.36$\;\;$[39.65, 45.92]$   &  $79.57$\;\;$[74.69, 84.47]$ \\
    4      & Tournament &  Fashion   &   SU(4)    &  $86.05$\;\;$[84.92, 87.17]$   &  $58.92$\;\;$[55.72, 62.15]$   &  $51.27$\;\;$[48.55, 54.13]$   &  $82.79$\;\;$[77.27, 88.31]$ \\
    \midrule
    5      & Tournament &  Fashion   &    CNN7    &  $74.35$\;\;$[72.75, 75.94]$   &  $54.18$\;\;$[51.03, 57.67]$   &  $41.24$\;\;$[38.23, 44.40]$   &  $79.07$\;\;$[73.79, 85.36]$ \\
    5      & Tournament &  Fashion   &    CNN8    &  $75.80$\;\;$[74.37, 77.05]$   &  $55.09$\;\;$[51.75, 58.73]$   &  $42.75$\;\;$[39.70, 46.29]$   &  $80.05$\;\;$[75.09, 86.20]$ \\
    5      & Tournament &  Fashion   &   SEL-Z    &  $70.72$\;\;$[66.85, 74.22]$   &  $41.93$\;\;$[39.08, 45.07]$   &  $30.48$\;\;$[28.77, 32.49]$   &  $66.36$\;\;$[61.08, 71.77]$ \\
    5      & Tournament &  Fashion   &   SEL-X    &  $59.38$\;\;$[58.41, 60.50]$   &  $32.96$\;\;$[31.61, 34.23]$   &  $19.81$\;\;$[18.99, 20.63]$   &  $64.34$\;\;$[60.99, 67.53]$ \\
    5      & Tournament &  Fashion   &   SO(4)    &  $70.11$\;\;$[69.05, 71.18]$   &  $48.81$\;\;$[46.62, 51.47]$   &  $34.97$\;\;$[33.25, 37.39]$   &  $78.50$\;\;$[73.59, 84.27]$ \\
    5      & Tournament &  Fashion   &   SU(4)    &  $78.19$\;\;$[77.09, 78.98]$   &  $56.59$\;\;$[53.13, 60.16]$   &  $45.14$\;\;$[42.36, 48.04]$   &  $80.94$\;\;$[76.92, 85.80]$ \\
    \midrule
    6      & Tournament &  Fashion   &    CNN7    &  $70.62$\;\;$[69.62, 72.05]$   &  $47.96$\;\;$[43.46, 55.48]$   &  $34.74$\;\;$[31.13, 41.09]$   &  $74.12$\;\;$[70.50, 79.28]$ \\
    6      & Tournament &  Fashion   &    CNN8    &  $71.03$\;\;$[69.67, 73.19]$   &  $49.17$\;\;$[45.09, 56.48]$   &  $36.03$\;\;$[32.41, 42.51]$   &  $74.63$\;\;$[70.85, 79.77]$ \\
    6      & Tournament &  Fashion   &   SEL-Z    &  $65.58$\;\;$[63.42, 67.74]$   &  $36.07$\;\;$[32.93, 41.67]$   &  $24.22$\;\;$[21.29, 28.91]$   &  $59.57$\;\;$[54.55, 66.03]$ \\
    6      & Tournament &  Fashion   &   SEL-X    &  $62.59$\;\;$[62.47, 62.69]$   &  $21.49$\;\;$[21.07, 22.02]$   &  $13.51$\;\;$[13.24, 13.87]$   &  $49.09$\;\;$[43.22, 54.97]$ \\
    6      & Tournament &  Fashion   &   SO(4)    &  $66.85$\;\;$[66.00, 67.70]$   &  $41.18$\;\;$[38.05, 46.24]$   &  $28.00$\;\;$[25.51, 31.68]$   &  $71.87$\;\;$[68.08, 77.19]$ \\
    6      & Tournament &  Fashion   &   SU(4)    &  $72.25$\;\;$[70.08, 74.31]$   &  $50.80$\;\;$[45.86, 58.23]$   &  $37.84$\;\;$[33.45, 44.41]$   &  $76.27$\;\;$[72.57, 81.10]$ \\

            \bottomrule
    \end{tabular}\medskip
    \label{tab:full_t}
\end{table*}

\begin{table*}[t]
\scriptsize
    \centering
    \caption{(Continued) Expanded metrics aggregated by class count, encoding, dataset, and circuit. The table reports resolvability ratio $R$, shot-level resolvable accuracy $A_R^{\mathrm{shot}}$, raw bitstring shot accuracy $A_b^{\mathrm{shot}}$, and continuous top-1 simulation accuracy $T$. The first three metrics evaluate discrete measurement outcomes: $R$ measures how often a shot yields a valid prediction, $A_R^{\mathrm{shot}}$ measures correctness conditional on validity, and $A_b^{\mathrm{shot}}$ counts unresolvable shots as incorrect. The metric $T$ reports accuracy under continuous pre-measurement decoding. Values are means with 95\% bootstrap confidence intervals $[CI^{\downarrow},CI^{\uparrow}]$ over matched experimental configurations. Larger values are better for all metrics.}
    \vskip .15in
    \centering
    \setlength\tabcolsep{2mm}
    \begin{tabular}{cccccccc}
        \toprule
        K & Method & Dataset & Circuit &  $R$ \;\;[$CI^{\downarrow}$, $CI^{\uparrow}$] & 
                     $A_R^{\mathrm{shot}}$ [$CI^{\downarrow}$, $CI^{\uparrow}$]\% & 
                     $A_b^{\mathrm{shot}}$ [$CI^{\downarrow}$, $CI^{\uparrow}$]\% & 
                     $T$\;\;[$CI^{\downarrow}$, $CI^{\uparrow}$]\% \\
        \midrule    
    3      &  One-hot   &   Digits   &    CNN8    &  $58.77$\;\;$[54.00, 62.84]$   &  $61.76$\;\;$[57.14, 67.34]$   &  $41.08$\;\;$[35.55, 46.66]$   &  $75.35$\;\;$[69.98, 80.47]$ \\
    3      &  One-hot   &   Digits   &   SEL-Z    &  $63.18$\;\;$[61.01, 66.36]$   &  $62.37$\;\;$[56.69, 68.14]$   &  $41.76$\;\;$[36.45, 46.57]$   &  $74.81$\;\;$[69.15, 79.59]$ \\
    3      &  One-hot   &   Digits   &   SEL-X    &  $57.51$\;\;$[52.75, 62.00]$   &  $61.97$\;\;$[57.67, 67.59]$   &  $40.30$\;\;$[34.33, 46.41]$   &  $74.99$\;\;$[69.17, 80.02]$ \\
    3      &  One-hot   &   Digits   &   SO(4)    &  $52.47$\;\;$[48.62, 55.62]$   &  $53.38$\;\;$[49.89, 56.87]$   &  $28.87$\;\;$[27.45, 30.10]$   &  $67.88$\;\;$[63.74, 71.38]$ \\
    3      &  One-hot   &   Digits   &   SU(4)    &  $57.54$\;\;$[54.36, 61.32]$   &  $61.60$\;\;$[56.46, 67.91]$   &  $40.29$\;\;$[35.58, 45.57]$   &  $75.68$\;\;$[71.05, 80.46]$ \\
\midrule
    4      &  One-hot   &   Digits   &    CNN7    &  $44.05$\;\;$[41.13, 46.47]$   &  $66.67$\;\;$[62.19, 70.75]$   &  $32.76$\;\;$[29.32, 36.20]$   &  $88.76$\;\;$[85.75, 90.86]$ \\
    4      &  One-hot   &   Digits   &    CNN8    &  $47.93$\;\;$[44.63, 51.54]$   &  $67.94$\;\;$[62.23, 72.29]$   &  $35.67$\;\;$[30.82, 40.01]$   &  $89.25$\;\;$[86.46, 91.36]$ \\
    4      &  One-hot   &   Digits   &   SEL-Z    &  $25.74$\;\;$[20.41, 31.08]$   &  $49.85$\;\;$[47.17, 52.54]$   &  $16.88$\;\;$[14.13, 19.62]$   &  $69.62$\;\;$[65.12, 73.40]$ \\
    4      &  One-hot   &   Digits   &   SEL-X    &  $28.33$\;\;$[25.02, 31.63]$   &  $51.46$\;\;$[45.91, 57.01]$   &  $17.17$\;\;$[13.53, 20.81]$   &  $80.21$\;\;$[71.96, 86.64]$ \\
    4      &  One-hot   &   Digits   &   SO(4)    &  $39.83$\;\;$[36.53, 42.85]$   &  $57.49$\;\;$[52.94, 61.15]$   &  $25.20$\;\;$[21.68, 27.97]$   &  $85.19$\;\;$[80.06, 88.69]$ \\
    4      &  One-hot   &   Digits   &   SU(4)    &  $49.81$\;\;$[45.88, 53.47]$   &  $68.10$\;\;$[62.87, 72.24]$   &  $36.78$\;\;$[31.63, 41.38]$   &  $89.52$\;\;$[86.44, 91.63]$ \\
\midrule
    5      &  One-hot   &   Digits   &    CNN7    &  $26.99$\;\;$[22.64, 30.96]$   &  $56.73$\;\;$[52.43, 60.32]$   &  $18.68$\;\;$[15.00, 21.51]$   &  $80.06$\;\;$[76.99, 82.21]$ \\
    5      &  One-hot   &   Digits   &    CNN8    &  $29.48$\;\;$[27.97, 30.91]$   &  $58.47$\;\;$[55.05, 61.25]$   &  $20.23$\;\;$[18.68, 21.71]$   &  $82.44$\;\;$[80.24, 84.18]$ \\
    5      &  One-hot   &   Digits   &   SEL-Z    &  $15.64$\;\;$[14.97, 16.36]$   &  $36.13$\;\;$[32.43, 40.46]$   &    $6.86$\;\;$[5.70, 8.04]$    &  $56.25$\;\;$[53.86, 60.43]$ \\
    5      &  One-hot   &   Digits   &   SEL-X    &  $15.95$\;\;$[14.54, 17.61]$   &  $36.97$\;\;$[31.95, 42.43]$   &    $6.98$\;\;$[4.91, 9.26]$    &  $65.60$\;\;$[61.18, 70.21]$ \\
    5      &  One-hot   &   Digits   &   SO(4)    &  $23.15$\;\;$[20.95, 25.48]$   &  $48.26$\;\;$[45.49, 51.66]$   &  $13.35$\;\;$[11.37, 15.62]$   &  $77.40$\;\;$[75.05, 79.49]$ \\
    5      &  One-hot   &   Digits   &   SU(4)    &  $31.77$\;\;$[27.57, 36.31]$   &  $61.34$\;\;$[57.14, 64.87]$   &  $23.09$\;\;$[19.03, 27.06]$   &  $84.17$\;\;$[82.43, 85.47]$ \\
\midrule
    6      &  One-hot   &   Digits   &    CNN7    &  $16.80$\;\;$[15.38, 17.89]$   &  $48.08$\;\;$[42.70, 51.18]$   &   $9.80$\;\;$[7.84, 10.86]$    &  $75.97$\;\;$[70.49, 80.53]$ \\
    6      &  One-hot   &   Digits   &    CNN8    &  $18.01$\;\;$[17.00, 19.02]$   &  $49.93$\;\;$[45.74, 53.53]$   &   $11.76$\;\;$[9.89, 13.45]$   &  $77.30$\;\;$[73.56, 82.01]$ \\
    6      &  One-hot   &   Digits   &   SEL-Z    &  $12.32$\;\;$[11.93, 12.74]$   &  $30.70$\;\;$[28.07, 33.49]$   &    $3.84$\;\;$[3.41, 4.34]$    &  $45.56$\;\;$[39.67, 51.09]$ \\
    6      &  One-hot   &   Digits   &   SEL-X    &  $10.63$\;\;$[10.47, 10.80]$   &  $23.08$\;\;$[21.71, 24.26]$   &    $2.10$\;\;$[1.98, 2.23]$    &  $46.46$\;\;$[45.64, 47.32]$ \\
    6      &  One-hot   &   Digits   &   SO(4)    &  $15.32$\;\;$[14.51, 16.10]$   &  $41.07$\;\;$[37.44, 44.40]$   &    $7.35$\;\;$[6.22, 8.48]$    &  $73.25$\;\;$[68.14, 77.18]$ \\
    6      &  One-hot   &   Digits   &   SU(4)    &  $20.77$\;\;$[18.70, 22.94]$   &  $55.76$\;\;$[51.32, 59.10]$   &  $14.42$\;\;$[11.67, 16.78]$   &  $81.05$\;\;$[78.28, 83.39]$ \\
\midrule
    3      &  One-hot   &  Fashion   &    CNN7    &  $55.57$\;\;$[35.44, 73.00]$   &  $59.93$\;\;$[50.68, 68.73]$   &  $44.10$\;\;$[33.55, 54.65]$   &  $73.86$\;\;$[63.31, 83.24]$ \\
    3      &  One-hot   &  Fashion   &    CNN8    &  $54.47$\;\;$[34.11, 72.93]$   &  $59.85$\;\;$[50.68, 68.56]$   &  $43.53$\;\;$[32.70, 54.37]$   &  $73.55$\;\;$[62.44, 83.33]$ \\
    3      &  One-hot   &  Fashion   &   SEL-Z    &  $68.13$\;\;$[55.30, 80.95]$   &  $61.84$\;\;$[54.71, 68.41]$   &  $44.22$\;\;$[31.63, 56.81]$   &  $74.66$\;\;$[65.72, 82.85]$ \\
    3      &  One-hot   &  Fashion   &   SEL-X    &  $55.59$\;\;$[34.90, 72.73]$   &  $61.16$\;\;$[53.16, 68.67]$   &  $42.80$\;\;$[31.70, 53.91]$   &  $73.31$\;\;$[62.79, 82.61]$ \\
    3      &  One-hot   &  Fashion   &   SO(4)    &  $48.83$\;\;$[43.40, 54.26]$   &  $60.12$\;\;$[54.40, 64.56]$   &  $31.06$\;\;$[26.12, 36.00]$   &  $74.93$\;\;$[65.88, 83.41]$ \\
    3      &  One-hot   &  Fashion   &   SU(4)    &  $54.85$\;\;$[34.95, 73.55]$   &  $60.14$\;\;$[50.82, 68.70]$   &  $43.83$\;\;$[32.65, 55.00]$   &  $75.05$\;\;$[66.61, 83.39]$ \\
\midrule
    4      &  One-hot   &  Fashion   &    CNN7    &  $44.05$\;\;$[38.19, 49.43]$   &  $59.66$\;\;$[55.82, 63.50]$   &  $30.15$\;\;$[26.75, 33.54]$   &  $82.19$\;\;$[77.34, 87.16]$ \\
    4      &  One-hot   &  Fashion   &    CNN8    &  $43.59$\;\;$[35.06, 51.82]$   &  $60.58$\;\;$[56.23, 64.93]$   &  $31.44$\;\;$[26.65, 36.22]$   &  $82.88$\;\;$[78.40, 87.36]$ \\
    4      &  One-hot   &  Fashion   &   SEL-Z    &  $28.55$\;\;$[23.06, 34.93]$   &  $47.07$\;\;$[44.87, 49.28]$   &  $15.81$\;\;$[12.84, 19.94]$   &  $67.98$\;\;$[61.32, 74.49]$ \\
    4      &  One-hot   &  Fashion   &   SEL-X    &  $27.31$\;\;$[25.27, 28.97]$   &  $48.19$\;\;$[46.04, 51.34]$   &  $15.47$\;\;$[13.64, 17.38]$   &  $78.38$\;\;$[73.06, 83.40]$ \\
    4      &  One-hot   &  Fashion   &   SO(4)    &  $36.54$\;\;$[30.18, 44.25]$   &  $53.89$\;\;$[49.51, 58.97]$   &  $23.16$\;\;$[19.37, 28.89]$   &  $80.68$\;\;$[75.75, 85.61]$ \\
    4      &  One-hot   &  Fashion   &   SU(4)    &  $48.09$\;\;$[40.73, 55.17]$   &  $61.54$\;\;$[57.32, 65.77]$   &  $34.08$\;\;$[29.12, 39.04]$   &  $82.78$\;\;$[78.25, 87.30]$ \\
\midrule
    5      &  One-hot   &  Fashion   &    CNN7    &  $28.60$\;\;$[25.64, 31.10]$   &  $55.75$\;\;$[52.45, 58.89]$   &  $19.98$\;\;$[18.38, 22.05]$   &  $79.12$\;\;$[75.77, 83.95]$ \\
    5      &  One-hot   &  Fashion   &    CNN8    &  $29.39$\;\;$[25.44, 33.79]$   &  $57.60$\;\;$[54.25, 61.99]$   &  $21.85$\;\;$[19.08, 25.69]$   &  $79.27$\;\;$[75.34, 84.78]$ \\
    5      &  One-hot   &  Fashion   &   SEL-Z    &  $18.67$\;\;$[16.04, 21.63]$   &  $40.08$\;\;$[33.74, 46.41]$   &   $10.10$\;\;$[6.76, 13.45]$   &  $65.38$\;\;$[59.49, 71.29]$ \\
    5      &  One-hot   &  Fashion   &   SEL-X    &  $17.25$\;\;$[16.22, 18.27]$   &  $40.37$\;\;$[35.79, 43.67]$   &    $7.67$\;\;$[6.44, 8.93]$    &  $66.86$\;\;$[62.17, 71.85]$ \\
    5      &  One-hot   &  Fashion   &   SO(4)    &  $25.58$\;\;$[24.16, 26.98]$   &  $52.16$\;\;$[50.09, 54.31]$   &  $16.30$\;\;$[14.75, 17.59]$   &  $77.24$\;\;$[73.76, 81.58]$ \\
    5      &  One-hot   &  Fashion   &   SU(4)    &  $30.96$\;\;$[26.69, 36.31]$   &  $58.73$\;\;$[54.98, 63.43]$   &  $23.63$\;\;$[19.93, 28.23]$   &  $80.87$\;\;$[77.26, 86.11]$ \\
\midrule
    6      &  One-hot   &  Fashion   &    CNN7    &  $16.11$\;\;$[15.05, 17.31]$   &  $45.44$\;\;$[42.12, 50.07]$   &   $9.39$\;\;$[7.85, 11.38]$    &  $72.51$\;\;$[69.08, 78.65]$ \\
    6      &  One-hot   &  Fashion   &    CNN8    &  $18.18$\;\;$[16.19, 21.30]$   &  $49.23$\;\;$[43.40, 55.67]$   &   $12.00$\;\;$[9.23, 15.37]$   &  $74.26$\;\;$[70.44, 79.74]$ \\
    6      &  One-hot   &  Fashion   &   SEL-Z    &  $12.25$\;\;$[11.22, 13.41]$   &  $31.56$\;\;$[27.92, 34.80]$   &    $3.98$\;\;$[3.04, 4.95]$    &  $53.15$\;\;$[47.28, 60.88]$ \\
    6      &  One-hot   &  Fashion   &   SEL-X    &  $10.83$\;\;$[10.68, 11.01]$   &  $25.00$\;\;$[23.78, 26.23]$   &    $2.43$\;\;$[2.24, 2.63]$    &  $51.57$\;\;$[48.19, 56.75]$ \\
    6      &  One-hot   &  Fashion   &   SO(4)    &  $15.40$\;\;$[13.95, 16.83]$   &  $42.29$\;\;$[39.83, 45.10]$   &    $7.62$\;\;$[6.61, 8.85]$    &  $71.15$\;\;$[67.03, 76.69]$ \\
    6      &  One-hot   &  Fashion   &   SU(4)    &  $20.26$\;\;$[18.41, 22.50]$   &  $52.87$\;\;$[48.84, 59.16]$   &  $14.27$\;\;$[12.54, 17.15]$   &  $75.61$\;\;$[72.20, 80.92]$ \\
            \bottomrule
    \end{tabular}\medskip
    \label{tab:full_o}
\end{table*}

\begin{table*}[t]
\scriptsize
    \centering
    \caption{(Continued) Expanded metrics aggregated by class count, encoding, dataset, and circuit. The table reports resolvability ratio $R$, shot-level resolvable accuracy $A_R^{\mathrm{shot}}$, raw bitstring shot accuracy $A_b^{\mathrm{shot}}$, and continuous top-1 simulation accuracy $T$. The first three metrics evaluate discrete measurement outcomes: $R$ measures how often a shot yields a valid prediction, $A_R^{\mathrm{shot}}$ measures correctness conditional on validity, and $A_b^{\mathrm{shot}}$ counts unresolvable shots as incorrect. The metric $T$ reports accuracy under continuous pre-measurement decoding. Values are means with 95\% bootstrap confidence intervals $[CI^{\downarrow},CI^{\uparrow}]$ over matched experimental configurations. Larger values are better for all metrics.}
    \vskip .15in
    \centering
    \setlength\tabcolsep{2mm}
    \begin{tabular}{cccccccc}
        \toprule
        K & Method & Dataset & Circuit &  $R$ \;\;[$CI^{\downarrow}$, $CI^{\uparrow}$] & 
                     $A_R^{\mathrm{shot}}$ [$CI^{\downarrow}$, $CI^{\uparrow}$]\% & 
                     $A_b^{\mathrm{shot}}$ [$CI^{\downarrow}$, $CI^{\uparrow}$]\% & 
                     $T$\;\;[$CI^{\downarrow}$, $CI^{\uparrow}$]\% \\
        \midrule    
    3      &  One-hot   &   Digits   &    CNN7    &  $56.91$\;\;$[53.91, 60.08]$   &  $62.76$\;\;$[57.76, 68.83]$   &  $40.59$\;\;$[35.71, 45.37]$   &  $75.53$\;\;$[70.82, 80.18]$ \\
    3      &   Binary   &   Digits   &    CNN7    &  $93.35$\;\;$[91.59, 94.81]$   &  $56.52$\;\;$[51.86, 61.46]$   &  $52.89$\;\;$[49.15, 56.74]$   &  $75.29$\;\;$[70.99, 81.19]$ \\
    3      &   Binary   &   Digits   &    CNN8    &  $93.55$\;\;$[90.85, 95.78]$   &  $56.34$\;\;$[51.77, 61.07]$   &  $52.83$\;\;$[49.04, 56.25]$   &  $73.52$\;\;$[68.33, 79.99]$ \\
    3      &   Binary   &   Digits   &   SEL-Z    &  $88.18$\;\;$[84.37, 91.49]$   &  $53.94$\;\;$[51.40, 56.24]$   &  $47.98$\;\;$[46.30, 49.76]$   &  $73.32$\;\;$[69.38, 76.69]$ \\
    3      &   Binary   &   Digits   &   SEL-X    &  $93.12$\;\;$[90.28, 94.88]$   &  $55.84$\;\;$[51.10, 59.96]$   &  $52.13$\;\;$[48.44, 54.71]$   &  $74.14$\;\;$[69.22, 80.38]$ \\
    3      &   Binary   &   Digits   &   SO(4)    &  $91.42$\;\;$[89.27, 93.06]$   &  $48.04$\;\;$[46.03, 50.18]$   &  $44.01$\;\;$[41.42, 46.60]$   &  $64.91$\;\;$[60.13, 69.69]$ \\
    3      &   Binary   &   Digits   &   SU(4)    &  $93.22$\;\;$[89.77, 95.74]$   &  $56.30$\;\;$[51.90, 60.82]$   &  $52.64$\;\;$[49.54, 55.49]$   &  $72.73$\;\;$[66.67, 79.80]$ \\
    \midrule
    4      &   Binary   &   Digits   &    CNN7    & $100.0$\;\;$[100.0, 100.0]$ &  $54.20$\;\;$[50.85, 57.37]$   &  $54.20$\;\;$[50.85, 57.37]$   &  $89.03$\;\;$[86.26, 90.97]$ \\
    4      &   Binary   &   Digits   &    CNN8    & $100.0$\;\;$[100.0, 100.0]$ &  $54.61$\;\;$[51.99, 56.32]$   &  $54.61$\;\;$[51.99, 56.32]$   &  $90.02$\;\;$[87.14, 91.62]$ \\
    4      &   Binary   &   Digits   &   SEL-Z    & $100.0$\;\;$[100.0, 100.0]$ &  $35.58$\;\;$[34.43, 37.19]$   &  $35.58$\;\;$[34.43, 37.19]$   &  $66.42$\;\;$[63.11, 69.55]$ \\
    4      &   Binary   &   Digits   &   SEL-X    & $100.0$\;\;$[100.0, 100.0]$ &  $45.60$\;\;$[43.68, 48.07]$   &  $45.60$\;\;$[43.68, 48.07]$   &  $81.59$\;\;$[77.40, 85.78]$ \\
    4      &   Binary   &   Digits   &   SO(4)    & $100.0$\;\;$[100.0, 100.0]$ &  $47.81$\;\;$[45.87, 49.01]$   &  $47.81$\;\;$[45.87, 49.01]$   &  $85.25$\;\;$[81.80, 88.06]$ \\
    4      &   Binary   &   Digits   &   SU(4)    & $100.0$\;\;$[100.0, 100.0]$ &  $56.42$\;\;$[52.72, 60.57]$   &  $56.42$\;\;$[52.72, 60.57]$   &  $90.81$\;\;$[88.39, 92.47]$ \\
    \midrule
    5      &   Binary   &   Digits   &    CNN7    &  $80.95$\;\;$[79.41, 82.50]$   &  $44.47$\;\;$[41.48, 47.16]$   &  $36.45$\;\;$[33.63, 38.83]$   &  $78.83$\;\;$[75.15, 81.37]$ \\
    5      &   Binary   &   Digits   &    CNN8    &  $82.02$\;\;$[80.75, 83.28]$   &  $47.36$\;\;$[44.80, 50.15]$   &  $39.39$\;\;$[36.85, 42.12]$   &  $81.88$\;\;$[78.90, 84.36]$ \\
    5      &   Binary   &   Digits   &   SEL-Z    &  $74.57$\;\;$[73.88, 75.26]$   &  $29.81$\;\;$[28.97, 30.65]$   &  $22.56$\;\;$[21.94, 23.09]$   &  $57.17$\;\;$[52.07, 61.31]$ \\
    5      &   Binary   &   Digits   &   SEL-X    &  $70.63$\;\;$[68.31, 73.01]$   &  $28.33$\;\;$[27.23, 30.05]$   &  $20.37$\;\;$[19.14, 21.75]$   &  $53.53$\;\;$[51.81, 55.25]$ \\
    5      &   Binary   &   Digits   &   SO(4)    &  $79.90$\;\;$[78.50, 81.09]$   &  $38.95$\;\;$[36.93, 40.96]$   &  $31.53$\;\;$[29.75, 33.30]$   &  $77.59$\;\;$[73.13, 82.05]$ \\
    5      &   Binary   &   Digits   &   SU(4)    &  $82.71$\;\;$[81.90, 83.52]$   &  $47.72$\;\;$[44.76, 49.99]$   &  $40.00$\;\;$[37.45, 41.96]$   &  $83.18$\;\;$[80.35, 85.19]$ \\
    \midrule
    6      &   Binary   &   Digits   &    CNN7    &  $85.31$\;\;$[84.16, 86.48]$   &  $34.94$\;\;$[32.48, 37.89]$   &  $30.14$\;\;$[27.69, 33.07]$   &  $68.92$\;\;$[64.18, 73.85]$ \\
    6      &   Binary   &   Digits   &    CNN8    &  $84.75$\;\;$[83.06, 86.06]$   &  $38.21$\;\;$[34.33, 42.10]$   &  $32.73$\;\;$[29.00, 36.52]$   &  $75.11$\;\;$[71.30, 78.92]$ \\
    6      &   Binary   &   Digits   &   SEL-Z    &  $81.88$\;\;$[80.75, 83.64]$   &  $22.50$\;\;$[21.53, 23.64]$   &  $18.59$\;\;$[17.59, 19.57]$   &  $42.68$\;\;$[38.58, 47.20]$ \\
    6      &   Binary   &   Digits   &   SEL-X    &  $78.16$\;\;$[77.00, 79.32]$   &  $21.45$\;\;$[20.43, 22.38]$   &  $16.91$\;\;$[16.30, 17.45]$   &  $40.55$\;\;$[37.81, 43.57]$ \\
    6      &   Binary   &   Digits   &   SO(4)    &  $83.47$\;\;$[83.13, 83.81]$   &  $30.11$\;\;$[28.72, 31.88]$   &  $25.38$\;\;$[24.21, 26.82]$   &  $64.99$\;\;$[61.15, 69.41]$ \\
    6      &   Binary   &   Digits   &   SU(4)    &  $85.59$\;\;$[84.51, 86.82]$   &  $38.89$\;\;$[36.54, 41.50]$   &  $33.63$\;\;$[31.61, 35.90]$   &  $78.04$\;\;$[75.43, 80.73]$ \\
    \midrule
    3      &   Binary   &  Fashion   &    CNN7    &  $91.32$\;\;$[88.22, 95.18]$   &  $57.14$\;\;$[49.33, 64.96]$   &  $52.88$\;\;$[46.25, 59.87]$   &  $76.90$\;\;$[67.87, 85.93]$ \\
    3      &   Binary   &  Fashion   &    CNN8    &  $91.09$\;\;$[86.90, 95.70]$   &  $57.46$\;\;$[49.80, 65.12]$   &  $53.19$\;\;$[46.10, 60.63]$   &  $76.70$\;\;$[67.73, 85.67]$ \\
    3      &   Binary   &  Fashion   &   SEL-Z    &  $88.59$\;\;$[81.32, 94.31]$   &  $54.93$\;\;$[45.59, 64.05]$   &  $50.23$\;\;$[40.17, 59.31]$   &  $69.29$\;\;$[50.01, 84.07]$ \\
    3      &   Binary   &  Fashion   &   SEL-X    &  $91.29$\;\;$[87.76, 94.86]$   &  $56.86$\;\;$[49.09, 64.57]$   &  $52.67$\;\;$[45.70, 60.31]$   &  $77.08$\;\;$[68.06, 86.10]$ \\
    3      &   Binary   &  Fashion   &   SO(4)    &  $91.16$\;\;$[87.12, 95.24]$   &  $56.16$\;\;$[49.33, 62.72]$   &  $51.93$\;\;$[46.05, 57.65]$   &  $76.20$\;\;$[67.83, 84.27]$ \\
    3      &   Binary   &  Fashion   &   SU(4)    &  $90.85$\;\;$[84.84, 96.49]$   &  $56.93$\;\;$[48.47, 64.85]$   &  $52.69$\;\;$[44.40, 60.35]$   &  $75.36$\;\;$[66.75, 83.97]$ \\
    \midrule
    4      &   Binary   &  Fashion   &    CNN7    & $100.0$\;\;$[100.0, 100.0]$ &  $49.83$\;\;$[46.82, 52.84]$   &  $49.83$\;\;$[46.82, 52.84]$   &  $82.48$\;\;$[76.68, 88.28]$ \\
    4      &   Binary   &  Fashion   &    CNN8    & $100.0$\;\;$[100.0, 100.0]$ &  $49.95$\;\;$[47.14, 52.77]$   &  $49.95$\;\;$[47.14, 52.77]$   &  $83.26$\;\;$[77.64, 88.88]$ \\
    4      &   Binary   &  Fashion   &   SEL-Z    & $100.0$\;\;$[100.0, 100.0]$ &  $35.82$\;\;$[34.89, 36.70]$   &  $35.82$\;\;$[34.89, 36.70]$   &  $67.12$\;\;$[59.65, 74.59]$ \\
    4      &   Binary   &  Fashion   &   SEL-X    & $100.0$\;\;$[100.0, 100.0]$ &  $43.62$\;\;$[42.21, 45.17]$   &  $43.62$\;\;$[42.21, 45.17]$   &  $79.67$\;\;$[74.73, 84.60]$ \\
    4      &   Binary   &  Fashion   &   SO(4)    & $100.0$\;\;$[100.0, 100.0]$ &  $45.04$\;\;$[42.28, 48.46]$   &  $45.04$\;\;$[42.28, 48.46]$   &  $81.31$\;\;$[75.59, 87.04]$ \\
    4      &   Binary   &  Fashion   &   SU(4)    & $100.0$\;\;$[100.0, 100.0]$ &  $49.82$\;\;$[46.73, 52.76]$   &  $49.82$\;\;$[46.73, 52.76]$   &  $82.80$\;\;$[76.27, 89.34]$ \\
    \midrule
    5      &   Binary   &  Fashion   &    CNN7    &  $84.13$\;\;$[81.32, 87.14]$   &  $45.70$\;\;$[41.79, 49.38]$   &  $38.84$\;\;$[34.96, 41.93]$   &  $76.40$\;\;$[71.56, 81.97]$ \\
    5      &   Binary   &  Fashion   &    CNN8    &  $84.61$\;\;$[80.98, 88.39]$   &  $48.13$\;\;$[44.19, 52.43]$   &  $41.29$\;\;$[37.05, 45.06]$   &  $79.33$\;\;$[75.18, 84.20]$ \\
    5      &   Binary   &  Fashion   &   SEL-Z    &  $76.41$\;\;$[74.09, 78.20]$   &  $32.24$\;\;$[30.16, 33.99]$   &  $24.95$\;\;$[22.74, 26.68]$   &  $59.75$\;\;$[57.44, 62.07]$ \\
    5      &   Binary   &  Fashion   &   SEL-X    &  $77.32$\;\;$[76.15, 78.50]$   &  $33.39$\;\;$[30.92, 35.85]$   &  $26.12$\;\;$[23.98, 28.32]$   &  $63.35$\;\;$[57.36, 70.44]$ \\
    5      &   Binary   &  Fashion   &   SO(4)    &  $82.49$\;\;$[80.05, 85.68]$   &  $41.49$\;\;$[38.63, 43.61]$   &  $34.62$\;\;$[31.88, 36.90]$   &  $74.94$\;\;$[71.13, 79.34]$ \\
    5      &   Binary   &  Fashion   &   SU(4)    &  $85.43$\;\;$[81.98, 89.00]$   &  $48.81$\;\;$[43.65, 53.42]$   &  $42.27$\;\;$[37.08, 46.35]$   &  $81.43$\;\;$[77.44, 85.66]$ \\
    \midrule
    6      &   Binary   &  Fashion   &    CNN7    &  $85.97$\;\;$[84.53, 87.48]$   &  $34.08$\;\;$[31.84, 37.09]$   &  $29.57$\;\;$[27.39, 32.51]$   &  $66.97$\;\;$[62.47, 73.96]$ \\
    6      &   Binary   &  Fashion   &    CNN8    &  $86.09$\;\;$[84.24, 88.22]$   &  $36.24$\;\;$[33.86, 39.36]$   &  $31.58$\;\;$[28.92, 34.62]$   &  $69.83$\;\;$[65.13, 76.04]$ \\
    6      &   Binary   &  Fashion   &   SEL-Z    &  $83.03$\;\;$[81.67, 84.39]$   &  $24.70$\;\;$[23.36, 26.50]$   &  $20.71$\;\;$[19.39, 22.58]$   &  $50.44$\;\;$[46.38, 55.52]$ \\
    6      &   Binary   &  Fashion   &   SEL-X    &  $79.45$\;\;$[78.29, 80.82]$   &  $21.71$\;\;$[20.82, 22.70]$   &  $17.37$\;\;$[16.86, 18.10]$   &  $42.75$\;\;$[41.03, 44.41]$ \\
    6      &   Binary   &  Fashion   &   SO(4)    &  $83.45$\;\;$[82.85, 84.09]$   &  $31.49$\;\;$[29.05, 34.58]$   &  $26.55$\;\;$[24.48, 29.25]$   &  $65.30$\;\;$[58.28, 73.98]$ \\
    6      &   Binary   &  Fashion   &   SU(4)    &  $86.39$\;\;$[84.96, 87.90]$   &  $37.84$\;\;$[35.48, 41.68]$   &  $33.13$\;\;$[30.69, 37.01]$   &  $71.71$\;\;$[67.00, 78.79]$ \\
            \bottomrule
    \end{tabular}\medskip
    \label{tab:full_b}
\end{table*}

\begin{table*}[t]
\scriptsize
    \centering
    \caption{(Continued) Expanded metrics aggregated by class count, encoding, dataset, and circuit. The table reports resolvability ratio $R$, shot-level resolvable accuracy $A_R^{\mathrm{shot}}$, raw bitstring shot accuracy $A_b^{\mathrm{shot}}$, and continuous top-1 simulation accuracy $T$. The first three metrics evaluate discrete measurement outcomes: $R$ measures how often a shot yields a valid prediction, $A_R^{\mathrm{shot}}$ measures correctness conditional on validity, and $A_b^{\mathrm{shot}}$ counts unresolvable shots as incorrect. The metric $T$ reports accuracy under continuous pre-measurement decoding. Values are means with 95\% bootstrap confidence intervals $[CI^{\downarrow},CI^{\uparrow}]$ over matched experimental configurations. Larger values are better for all metrics.}
    \vskip .15in
    \centering
    \setlength\tabcolsep{2mm}
    \begin{tabular}{cccccccc}
        \toprule
        K & Method & Dataset & Circuit &  $R$ \;\;[$CI^{\downarrow}$, $CI^{\uparrow}$] & 
                     $A_R^{\mathrm{shot}}$ [$CI^{\downarrow}$, $CI^{\uparrow}$]\% & 
                     $A_b^{\mathrm{shot}}$ [$CI^{\downarrow}$, $CI^{\uparrow}$]\% & 
                     $T$\;\;[$CI^{\downarrow}$, $CI^{\uparrow}$]\% \\
        \midrule    
    3      &    Gray    &   Digits   &    CNN7    &  $92.36$\;\;$[89.34, 95.20]$   &  $55.52$\;\;$[51.32, 60.59]$   &  $51.90$\;\;$[47.78, 57.04]$   &  $75.46$\;\;$[71.29, 80.88]$ \\
    3      &    Gray    &   Digits   &    CNN8    &  $92.17$\;\;$[89.53, 94.80]$   &  $56.15$\;\;$[51.93, 61.02]$   &  $52.39$\;\;$[48.89, 57.19]$   &  $75.38$\;\;$[70.63, 80.63]$ \\
    3      &    Gray    &   Digits   &   SEL-Z    &  $88.50$\;\;$[87.18, 89.80]$   &  $55.64$\;\;$[50.93, 60.30]$   &  $49.69$\;\;$[45.23, 53.59]$   &  $74.64$\;\;$[68.75, 80.07]$ \\
    3      &    Gray    &   Digits   &   SEL-X    &  $89.76$\;\;$[86.78, 92.69]$   &  $55.96$\;\;$[51.92, 60.84]$   &  $50.84$\;\;$[47.06, 55.92]$   &  $75.78$\;\;$[70.85, 81.27]$ \\
    3      &    Gray    &   Digits   &   SO(4)    &  $89.46$\;\;$[86.93, 91.36]$   &  $47.65$\;\;$[45.19, 49.67]$   &  $42.90$\;\;$[41.06, 44.56]$   &  $67.90$\;\;$[63.51, 70.99]$ \\
    3      &    Gray    &   Digits   &   SU(4)    &  $92.23$\;\;$[88.66, 95.28]$   &  $55.48$\;\;$[50.89, 60.60]$   &  $51.89$\;\;$[47.39, 56.93]$   &  $74.78$\;\;$[70.72, 80.19]$ \\
\midrule
    4      &    Gray    &   Digits   &    CNN7    & $100.0$\;\;$[100.0, 100.0]$ &  $24.88$\;\;$[23.59, 26.08]$   &  $24.88$\;\;$[23.59, 26.08]$   &  $24.93$\;\;$[24.16, 26.05]$ \\
    4      &    Gray    &   Digits   &    CNN8    & $100.0$\;\;$[100.0, 100.0]$ &  $25.07$\;\;$[23.09, 26.59]$   &  $25.07$\;\;$[23.09, 26.59]$   &  $25.14$\;\;$[24.21, 26.24]$ \\
    4      &    Gray    &   Digits   &   SEL-Z    & $100.0$\;\;$[100.0, 100.0]$ &  $25.36$\;\;$[23.66, 27.27]$   &  $25.36$\;\;$[23.66, 27.27]$   &  $24.39$\;\;$[20.53, 28.24]$ \\
    4      &    Gray    &   Digits   &   SEL-X    & $100.0$\;\;$[100.0, 100.0]$ &  $24.81$\;\;$[23.56, 26.06]$   &  $24.81$\;\;$[23.56, 26.06]$   &  $25.46$\;\;$[23.94, 26.74]$ \\
    4      &    Gray    &   Digits   &   SO(4)    & $100.0$\;\;$[100.0, 100.0]$ &  $25.08$\;\;$[23.54, 26.41]$   &  $25.08$\;\;$[23.54, 26.41]$   &  $25.46$\;\;$[24.55, 26.45]$ \\
    4      &    Gray    &   Digits   &   SU(4)    & $100.0$\;\;$[100.0, 100.0]$ &  $25.13$\;\;$[23.77, 26.16]$   &  $25.13$\;\;$[23.77, 26.16]$   &  $25.45$\;\;$[25.00, 26.11]$ \\
\midrule
    5      &    Gray    &   Digits   &    CNN7    &  $79.86$\;\;$[78.05, 82.05]$   &  $43.91$\;\;$[41.52, 46.77]$   &  $35.60$\;\;$[33.18, 38.29]$   &  $79.71$\;\;$[77.55, 81.87]$ \\
    5      &    Gray    &   Digits   &    CNN8    &  $82.04$\;\;$[80.72, 83.93]$   &  $46.02$\;\;$[42.46, 50.05]$   &  $38.31$\;\;$[34.93, 42.42]$   &  $81.84$\;\;$[78.67, 84.98]$ \\
    5      &    Gray    &   Digits   &   SEL-Z    &  $73.31$\;\;$[71.60, 75.01]$   &  $29.07$\;\;$[27.39, 30.66]$   &  $21.68$\;\;$[20.20, 23.16]$   &  $54.32$\;\;$[47.74, 59.98]$ \\
    5      &    Gray    &   Digits   &   SEL-X    &  $72.42$\;\;$[68.36, 77.23]$   &  $31.84$\;\;$[28.62, 36.43]$   &  $23.63$\;\;$[19.95, 28.20]$   &  $59.47$\;\;$[53.42, 67.22]$ \\
    5      &    Gray    &   Digits   &   SO(4)    &  $79.23$\;\;$[78.20, 80.47]$   &  $38.65$\;\;$[35.86, 41.91]$   &  $31.16$\;\;$[28.59, 34.35]$   &  $75.23$\;\;$[70.91, 78.74]$ \\
    5      &    Gray    &   Digits   &   SU(4)    &  $83.61$\;\;$[82.38, 85.74]$   &  $49.09$\;\;$[45.50, 53.57]$   &  $41.68$\;\;$[38.09, 46.47]$   &  $83.71$\;\;$[81.44, 86.15]$ \\
\midrule
    6      &    Gray    &   Digits   &    CNN7    &  $85.46$\;\;$[84.21, 86.71]$   &  $36.16$\;\;$[33.77, 38.67]$   &  $31.22$\;\;$[29.10, 33.40]$   &  $72.55$\;\;$[69.08, 75.87]$ \\
    6      &    Gray    &   Digits   &    CNN8    &  $86.54$\;\;$[85.35, 87.73]$   &  $37.86$\;\;$[35.71, 40.01]$   &  $33.10$\;\;$[31.35, 34.84]$   &  $76.53$\;\;$[73.24, 79.81]$ \\
    6      &    Gray    &   Digits   &   SEL-Z    &  $81.79$\;\;$[79.99, 83.59]$   &  $23.56$\;\;$[22.38, 24.74]$   &  $19.39$\;\;$[18.63, 20.14]$   &  $46.47$\;\;$[42.09, 50.19]$ \\
    6      &    Gray    &   Digits   &   SEL-X    &  $78.55$\;\;$[77.42, 79.67]$   &  $21.26$\;\;$[19.93, 22.60]$   &  $16.88$\;\;$[15.79, 17.96]$   &  $35.53$\;\;$[31.83, 39.23]$ \\
    6      &    Gray    &   Digits   &   SO(4)    &  $83.30$\;\;$[82.73, 84.14]$   &  $29.83$\;\;$[28.32, 31.34]$   &  $25.18$\;\;$[24.03, 26.36]$   &  $62.54$\;\;$[59.38, 65.18]$ \\
    6      &    Gray    &   Digits   &   SU(4)    &  $86.64$\;\;$[85.48, 87.55]$   &  $39.69$\;\;$[37.50, 41.93]$   &  $34.75$\;\;$[32.52, 36.98]$   &  $78.36$\;\;$[75.04, 82.10]$ \\
\midrule
    3      &    Gray    &  Fashion   &    CNN7    &  $95.62$\;\;$[90.38, 99.04]$   &  $57.13$\;\;$[49.64, 64.57]$   &  $55.13$\;\;$[45.87, 63.97]$   &  $76.15$\;\;$[66.95, 85.35]$ \\
    3      &    Gray    &  Fashion   &    CNN8    &  $96.50$\;\;$[93.22, 98.92]$   &  $57.83$\;\;$[50.51, 65.24]$   &  $56.10$\;\;$[47.63, 64.57]$   &  $76.11$\;\;$[66.67, 85.54]$ \\
    3      &    Gray    &  Fashion   &   SEL-Z    &  $93.06$\;\;$[90.67, 95.31]$   &  $56.02$\;\;$[49.68, 62.45]$   &  $52.31$\;\;$[46.43, 57.98]$   &  $75.93$\;\;$[66.28, 85.97]$ \\
    3      &    Gray    &  Fashion   &   SEL-X    &  $95.74$\;\;$[93.65, 97.91]$   &  $57.12$\;\;$[50.68, 63.57]$   &  $54.99$\;\;$[47.94, 62.11]$   &  $76.59$\;\;$[67.49, 85.67]$ \\
    3      &    Gray    &  Fashion   &   SO(4)    &  $93.70$\;\;$[89.41, 97.56]$   &  $55.67$\;\;$[50.19, 61.16]$   &  $52.66$\;\;$[46.20, 58.71]$   &  $75.74$\;\;$[66.22, 85.39]$ \\
    3      &    Gray    &  Fashion   &   SU(4)    &  $96.03$\;\;$[90.88, 98.99]$   &  $57.61$\;\;$[49.89, 65.27]$   &  $55.79$\;\;$[46.59, 64.42]$   &  $76.44$\;\;$[67.09, 85.79]$ \\
\midrule
    4      &    Gray    &  Fashion   &    CNN7    & $100.0$\;\;$[100.0, 100.0]$ &  $24.99$\;\;$[23.44, 26.52]$   &  $24.99$\;\;$[23.44, 26.52]$   &  $26.52$\;\;$[23.84, 28.94]$ \\
    4      &    Gray    &  Fashion   &    CNN8    & $100.0$\;\;$[100.0, 100.0]$ &  $25.29$\;\;$[23.83, 26.75]$   &  $25.29$\;\;$[23.83, 26.75]$   &  $26.53$\;\;$[23.22, 29.34]$ \\
    4      &    Gray    &  Fashion   &   SEL-Z    & $100.0$\;\;$[100.0, 100.0]$ &  $25.26$\;\;$[24.12, 25.94]$   &  $25.26$\;\;$[24.12, 25.94]$   &  $27.74$\;\;$[24.87, 30.12]$ \\
    4      &    Gray    &  Fashion   &   SEL-X    & $100.0$\;\;$[100.0, 100.0]$ &  $25.50$\;\;$[24.43, 26.28]$   &  $25.50$\;\;$[24.43, 26.28]$   &  $26.58$\;\;$[23.75, 29.34]$ \\
    4      &    Gray    &  Fashion   &   SO(4)    & $100.0$\;\;$[100.0, 100.0]$ &  $25.14$\;\;$[23.93, 26.35]$   &  $25.14$\;\;$[23.93, 26.35]$   &  $26.33$\;\;$[22.50, 29.54]$ \\
    4      &    Gray    &  Fashion   &   SU(4)    & $100.0$\;\;$[100.0, 100.0]$ &  $25.16$\;\;$[23.69, 26.48]$   &  $25.16$\;\;$[23.69, 26.48]$   &  $26.75$\;\;$[22.85, 30.00]$ \\
\midrule
    5      &    Gray    &  Fashion   &    CNN7    &  $84.52$\;\;$[82.16, 88.20]$   &  $47.31$\;\;$[45.39, 49.72]$   &  $40.31$\;\;$[37.89, 42.74]$   &  $77.76$\;\;$[72.97, 83.44]$ \\
    5      &    Gray    &  Fashion   &    CNN8    &  $84.91$\;\;$[83.05, 87.68]$   &  $47.78$\;\;$[45.05, 51.25]$   &  $40.86$\;\;$[37.98, 43.90]$   &  $79.95$\;\;$[75.35, 85.38]$ \\
    5      &    Gray    &  Fashion   &   SEL-Z    &  $77.24$\;\;$[76.07, 78.29]$   &  $34.49$\;\;$[32.57, 36.89]$   &  $26.85$\;\;$[25.21, 29.07]$   &  $65.80$\;\;$[62.54, 70.57]$ \\
    5      &    Gray    &  Fashion   &   SEL-X    &  $75.94$\;\;$[73.30, 78.72]$   &  $33.83$\;\;$[30.98, 36.68]$   &  $26.07$\;\;$[24.04, 27.95]$   &  $64.08$\;\;$[59.70, 68.28]$ \\
    5      &    Gray    &  Fashion   &   SO(4)    &  $82.18$\;\;$[79.65, 85.58]$   &  $42.60$\;\;$[39.68, 45.52]$   &  $35.29$\;\;$[32.09, 38.48]$   &  $75.11$\;\;$[71.15, 81.13]$ \\
    5      &    Gray    &  Fashion   &   SU(4)    &  $85.75$\;\;$[83.76, 88.55]$   &  $49.74$\;\;$[46.74, 53.38]$   &  $43.03$\;\;$[39.80, 46.62]$   &  $81.47$\;\;$[76.42, 87.20]$ \\
\midrule
    6      &    Gray    &  Fashion   &    CNN7    &  $84.93$\;\;$[84.23, 85.82]$   &  $34.86$\;\;$[31.60, 38.47]$   &  $30.09$\;\;$[27.17, 33.37]$   &  $67.59$\;\;$[60.92, 75.55]$ \\
    6      &    Gray    &  Fashion   &    CNN8    &  $87.43$\;\;$[86.50, 88.49]$   &  $36.95$\;\;$[34.12, 40.80]$   &  $32.68$\;\;$[30.05, 36.11]$   &  $72.10$\;\;$[68.09, 77.88]$ \\
    6      &    Gray    &  Fashion   &   SEL-Z    &  $82.86$\;\;$[81.64, 84.08]$   &  $25.03$\;\;$[24.26, 25.93]$   &  $20.95$\;\;$[20.32, 21.58]$   &  $51.82$\;\;$[50.31, 53.82]$ \\
    6      &    Gray    &  Fashion   &   SEL-X    &  $78.72$\;\;$[76.41, 80.53]$   &  $22.85$\;\;$[21.64, 24.74]$   &  $18.17$\;\;$[17.08, 19.94]$   &  $43.50$\;\;$[36.44, 50.64]$ \\
    6      &    Gray    &  Fashion   &   SO(4)    &  $84.79$\;\;$[84.29, 85.30]$   &  $31.86$\;\;$[30.01, 33.94]$   &  $27.35$\;\;$[25.82, 29.20]$   &  $67.44$\;\;$[63.90, 72.62]$ \\
    6      &    Gray    &  Fashion   &   SU(4)    &  $86.74$\;\;$[86.15, 87.52]$   &  $38.17$\;\;$[35.32, 42.07]$   &  $33.54$\;\;$[30.87, 36.94]$   &  $75.01$\;\;$[70.99, 80.56]$ \\ 

            \bottomrule
    \end{tabular}\medskip
    \label{tab:full_g}
\end{table*}

\end{document}